\documentclass[twocolumn]{article}

\usepackage{geometry}
\usepackage[utf8]{inputenc}
\usepackage{hyperref}

\usepackage{cite}
\usepackage{amsmath,amssymb}
\usepackage{algorithmic}
\usepackage{graphicx}
\usepackage{textcomp}
\usepackage{multirow}
\usepackage{ulem}
\usepackage{subcaption}
\usepackage[ruled,norelsize]{algorithm2e}
\usepackage{nomencl}
\usepackage{xcolor}

\makeatletter
\newcommand{\removelatexerror}{\let\@latex@error\@gobble}
\makeatother

\usepackage{ftnxtra}
\usepackage{fnpos}

\usepackage{adjustbox}
\usepackage{multirow}
\usepackage{subcaption}
\usepackage{array}
\usepackage[font=small,skip=1pt]{caption}
\newcolumntype{C}[1]{>{\centering\let\newline\\\arraybackslash\hspace{0pt}}m{#1}}

\title{Intelligent Residential Energy Management System using Deep Reinforcement Learning}
\author{Alwyn Mathew 
\and Abhijit Roy \and Jimson Mathew}
\date{
	Indian Institute of Technology Patna \\ \texttt{\{alwyn.pcs16, abhijit.cs15, jimson\}@iitp.ac.in}\\[2ex]%
}

\begin{document}
	
\setlength{\abovedisplayskip}{0pt}
\setlength{\belowdisplayskip}{0pt}

\maketitle
	
\begin{abstract}
	The rising demand for electricity and its essential nature in today's world calls for intelligent home energy management (HEM) systems that can reduce energy usage. This involves scheduling of loads from peak hours of the day when energy consumption is at its highest to leaner off-peak periods of the day when energy consumption is relatively lower thereby reducing the system's peak load demand, which would consequently result in lesser energy bills, and improved load demand profile. This work introduces a novel way to develop a learning system that can learn from experience to shift loads from one time instance to another and achieve the goal of minimizing the aggregate peak load. This paper proposes a Deep Reinforcement Learning (DRL) model for demand response where the virtual agent learns the task like humans do. The agent gets feedback for every action it takes in the environment; these feedbacks will drive the agent to learn about the environment and take much smarter steps later in its learning stages. Our method outperformed the state of the art mixed integer linear programming (MILP) for load peak reduction. The authors have also designed an agent to learn to minimize both consumers' electricity bills and utilities' system peak load demand simultaneously. The proposed model was analyzed with loads from five different residential consumers; the proposed method increases the monthly savings of each consumer by reducing their electricity bill drastically along with minimizing the peak load on the system when time shiftable loads are handled by the proposed method.
	
\end{abstract}

\let\thefootnote\relax\footnote{© 2020 IEEE.  Personal use of this material is permitted.  Permission from IEEE must be obtained for all other uses, in any current or future media, including reprinting/republishing this material for advertising or promotional purposes, creating new collective works, for resale or redistribution to servers or lists, or reuse of any copyrighted component of this work in other works. This work was partially supported by the Scheme for Promotion of Academic and Research Collaboration (SPARC), MHRD, Government of India under Grant \# ID: P582.}

\section{Introduction}

Energy generated from power grids fuels the modern lifestyle. Per-user consumption of energy is ever increasing. People, nowadays, use a lot of modern appliances for their day to day chores. With technological advances, the invention of new appliances, and the ever-increasing interest of the new generations in the gadget market, investment in the household appliance market has increased manifold. With most of these appliances running on electricity, the rising electricity consumption also increases the load on power grids. People nowadays are willing to pay more for electricity rather than live without it. Household appliances in the US are responsible for approximately 42\% of the total energy consumption \cite{c1}. Given the high demand for electricity, efforts are being made continuously to improve smart grids with advanced research in power system and computer science. Rising energy requirement increases the load on power grids. Also, energy consumption follow specific trends that lead to disparity in demands from grids based on the time of the day, i.e, energy demand during particular periods can be higher than usual, whereas during other periods of the day energy requirements can be pretty low. During peak hours, the load on power grids increases drastically. To avert this problem, Demand Side Management (DSM) strategies are used. DSM strategies involve Demand Response (DR), energy efficiency, and conservation schemes.

\begin{figure*}
	\centering
	\includegraphics[width=1.5\columnwidth]{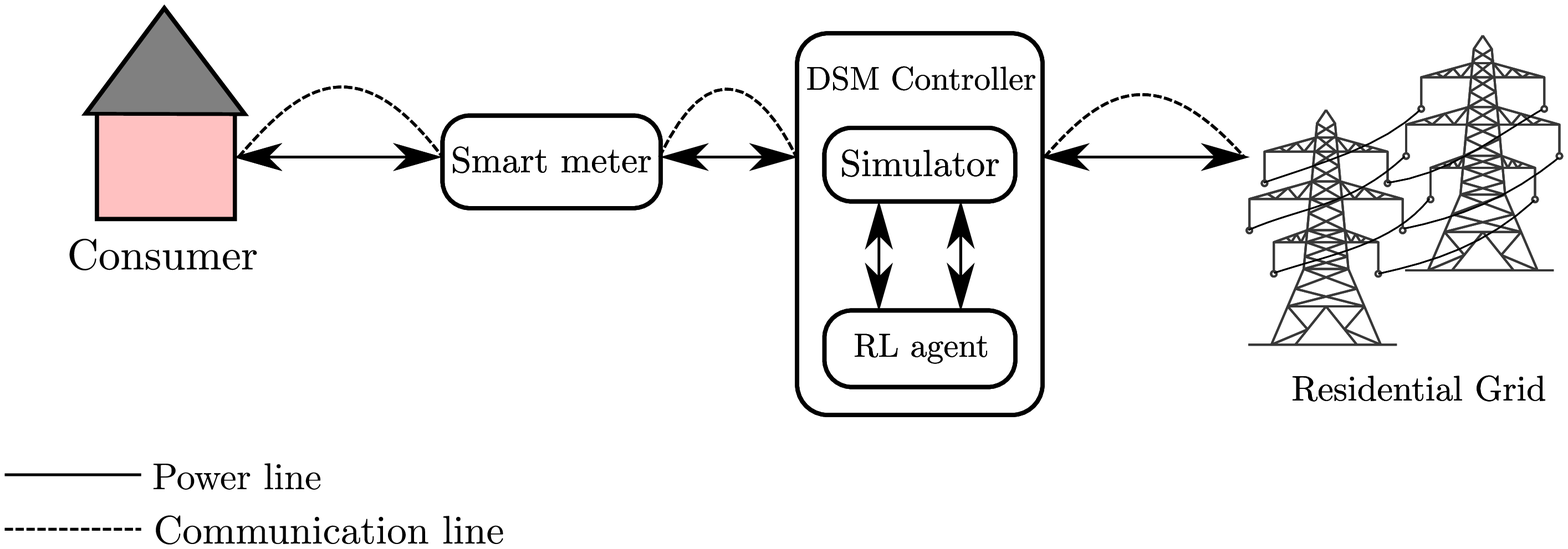}
	\caption{Proposed architecture of smart gird with RL-DSM at consumer end.}
	\label{fig:intro}
\end{figure*}

DR \cite{c2, c25} focuses on modifying consumers' demand on the benefit of reducing peak load on the grid and in turn, giving some incentives back to customers who participate in it. DR encourages consumers to use less power during peak hours of the day or shift their time of energy usage to off-peak hours of the day. Examples of DR may include storing energy during the off-peak periods from the grid and using the stored energy during the peak period. One could also make use of renewable sources of energy by storing these energies such as solar, wind, geothermal and biogas, which could be used during the peak periods of the day. The benefits of DSM are discussed in \cite{c3}. Home Energy Management (HEM) \cite{c27} is a part of DSM. HEM systems manage the usage of electricity in smart homes. DR is a crucial component of HEM systems. DR \cite{c26} is one of the techniques used in DSM. It involves scheduling loads on the timescale by moving high wattage loads to a different time to reduce the maximum load on the grid without changing the net energy consumed. It focuses on changing the ``when consumed" rather than changing the ``how much consumed".

Electricity generation involves several generating stations employing different generation technologies working in conjunction, thus making the process dynamic in its behavior. This translates to varying costs of electricity generation at any given point in time. This is where load shifting comes into play. Scheduling the load on the timescale that would reduce the overall peak demand on the grid and also saves electricity bills for the consumers. Most of the DR based optimization models are based on two broad categories: price-based and incentive-based. Price based optimization is discussed in the study conducted in this paper. The price-based models consider Time of Use (ToU) pricing, peak load pricing, critical peak pricing, and real-time pricing \cite{c28, c29}, which take into account the peak and off-peak tariffs. The varying tariffs of energy consumption during the entire duration of a day based on the aggregate load on the electrical systems act as a motivation for consumers to adjust their appliance usage to take advantage of the lower prices during specific periods. 
Incentive-Based DR programs are two types: Classical programs and Market-based programs. Classical programs include Direct Load Control and Interruptible programs, Market-based are Emergency DR, Demand Bidding, Capacity Market, and Ancillary services market \cite{albadi2007demand}. In \cite{caron2010incentive}, the authors proposed a pricing scheme for consumers with incentives to achieve a lower aggregate load profile. They also studied the load demand minimization possible with the amount of information that consumers share. In \cite{ghazvini2015multi} linear and nonlinear modeling for incentive-based DR for real power markets was proposed. System-level dispatch of demand response resources with a novel incentive-based demand response model was proposed by \cite{yu2017incentive}.
In \cite{aalami2010demand}, the authors proposes an Interruptible program, including penalties for customers in case they do not respond to load reduction. A real-time implementation of incentive-based DR programs with hardware for residential buildings is shown in \cite{caron2010incentive}. In \cite{zhong2012coupon} a novel DR program targeting small to medium size commercial, industrial,  and residential customers is proposed.

Reinforcement Learning (RL) \cite{c4} is an area of machine learning in computer science where a learning agent interacts with an environment and receives rewards as feedback of the interaction with the ultimate goal of maximizing the cumulative reward. To achieve this, RL agents attempt to come up with a policy mapping states to the best action at any given state, which would result in maximum expected future rewards. Well designed RL agents have displayed impeccable decision-making capabilities, such as Google's AlphaGo and OpenAI Five, in complex environments without the requirement of any prior domain knowledge. Deep Reinforcement Learning (DRL) \cite{c37} is a merger of deep learning and reinforcement learning where deep learning architectures such as neural networks are used with reinforcement learning algorithms like Q-learning, actor-critic, etc.  \cite{c5} discusses building DRL agents for playing Atari games and how DQN (Deep Q-Network) shows exceptional performance in playing the games.

The proposed work aim at applying deep reinforcement learning techniques to the scenario of load shifting and comparing the results obtained with that of the MILP \cite{c12,c15,c16} based methods. We also propose a smart grid architecture, as shown in Figure~\ref{fig:intro} where RL based DSM controller can be placed at the consumer end.   The grid end DSM with RL is an extension of this work, where raw loads come from residential microgrids instead of individual homes. An automated RL agent performing the task of load shifting would go a long way into optimizing load consumption by distributing loads from peak to off-peak periods, thus reducing the total load on the power grid during peak periods and reducing the energy bills of consumers. The main contributions of the proposed work are:

\begin{enumerate}
	\item Introduced Deep reinforcement learning in Demand Side
	Management (RL-DSM) for DR.
	\item Analyzed that the impact of a well-calculated reward system is crucial for Demand Side Management Reinforcement Learning models.
	\item The proposed reinforcement learning model surpassed traditional methods with a single objective by saving 6.04\% of the monthly utility bill.
	\item The proposed reinforcement learning model with multi-objective saved 11.66\% of the monthly utility bill, which shows the superior ability of the learning model over traditional methods for Demand Side Management.
\end{enumerate}

\section{Related Work}


Demand response optimization is extensively explored in literature. \cite{palensky2011demand} gives an overview of DSM with various types and the latest demonstration projects in DSM. \cite{c30} discusses a greedy iterative algorithm that enables users to schedule appliances. \cite{c31} presents linear programming based load scheduling algorithms. Mixed-integer linear programming model for optimal scheduling of appliances has been discussed in \cite{c12,c15,c16}. Heuristic-based scheduling algorithms that aim at cost minimization, user comfort maximization, and peak to average ratio minimization have been discussed in detail in \cite{c32}. A constrained multi-objective scheduling model for the purpose of optimizing utility and minimizing cost is proposed in \cite{c33}. A dynamic pricing model for energy consumption cost minimization and comfort maximization in the context of smart homes has been explored in \cite{c34}. A study of various control algorithms and architectures applied to DR was carried out in \cite{c19}.  \cite{d2019mapping} proposes a demand response cost minimization strategy with an air source heat pump along with a water thermal storage system in a building. \cite{guo2018machine} studies various energy demand prediction machine learning models like feed-forward neural network, support vector machine, and multiple linear regression. \cite{chen2019quantification} proposes a systematic method to quantify building electricity flexibility, which can be an assuring demand response resource, especially for large commercial buildings. \cite{alimohammadisagvand2018comparison}  studies the effect of demand response on energy consumption minimization with the heat generation system in small residential builds in a more frigid climate.

Increasing demand for energy has shifted the focus of the power industry to alternative renewable sources of energy for power generation. Integrating renewable sources of energy into HEM systems can be beneficial for both customers and power generation companies. Renewable sources of energy can fill in the excess demand for electricity by customers, thus moderating the peak loads on the grids. Aghaei \textit{et al.} carried out studies on DR using renewable sources of energy \cite{c18}. \cite{cui2018two} introduces a two-stage power-sharing framework. \cite{yaghmaee2016autonomous} talks about a cloud-based DSM method where consumer also has local power generation with batteries. \cite{c35} discusses the design for smart homes with the integration of renewable energy sources for peak load reduction and energy bill minimization. \cite{tushar2017demand} introduces a real-time decentralized DSM algorithm that takes advantage of energy storage systems (ESS), renewable energy, and regularizing charging/discharging. \cite{ma2018spectrum} proposes a method to improve DSM by optimizing power and spectrum allocation. \cite{li2017multiobjective} proposes a hierarchical day-ahead DSM model with renewable energy.  \cite{c36} discusses an Intelligent Residential Energy Management System (IREMS), which offers a model with in-house power generation using renewable sources of energy. 
\cite{li2018residential} proposes an algorithm to minimize consumer cost by facilitating energy buying and selling between utility and residential community. \cite{chiu2019pareto} offers a multi-objective to demand response program by reducing the energy cost of residential consumers and peak demand of the grid. \cite{rajasekhar2019collaborative} introduces collaborative energy trading and load scheduling using a game-theoretic approach and genetic algorithm. \cite{rajasekhar2015decentralized} propose a game-theoretic approach to minimize consumer energy cost and discomfort in a heterogeneous residential neighborhood. \cite{li2016real} reduces the overall cost of the system by optimizing the load scheduling and energy storage control simultaneously with Lyapunov optimization. \cite{arun2017intelligent} proposes an intelligent residential energy management system for residential buildings to reduce peak power demand and reducing prosumers' electrics bills. \cite{martinez2018optimizing} introduces an automated smart home energy management system using L-BFGS-B (Limited-memory Broyden–Fletcher–Goldfarb–Shanno) algorithm with time-of-use pricing to optimize the load schedule.


\cite{c17} introduces an RL model to meet the overall demand with the current load, and the next 24 hrs load predicted information by shifting loads. \cite{c20} formed a fully automated energy management system by decomposing rescheduling loads over device clusters. \cite{c21} proposes the Consumer Automated Energy Management System (CAES), an online learning model that estimates the influence of future energy prices and schedules device loads. \cite{c38} proposes a decentralized learning-based multi-agent residential DR for efficient usage of renewable sources in the grid. Lu \cite{c39} dynamic pricing DR algorithm formulated as an MDP to promote service providers profit, reduce costs for consumers, and attain efficiency in energy demand and supply. \cite{c40} introduced RL based scheduling of controllable load under a day-ahead pricing scheme.

This work encroaches into the territory of applying DRL to DR. Existing deep reinforcement learning-based model for DR, which came out recently are \cite{c22,c24}. \cite{c22} discusses an RL and DNN based algorithm modeled on top of real-time incentives. A complete review of a collection of all control algorithms and their assessment in DR strategies in the residential sector is given in \cite{c24}. It discusses machine learning-based predictive approaches and rule-based approaches.  Applying deep reinforcement learning models to the setting of DR can be a lucrative field of research. The proposed model draws inspiration from DRL agents used to play the game of Tetris. We have explored models of DRL automated agents playing the game of Tetris and have tried to look for areas of similarity in the formulation of the reward function. \cite{c13} discusses the design of a DQN model for playing Tetris and presents the reward system of an automated Tetris agent.

\begin{figure*}
	\centering
	\includegraphics[width=2\columnwidth]{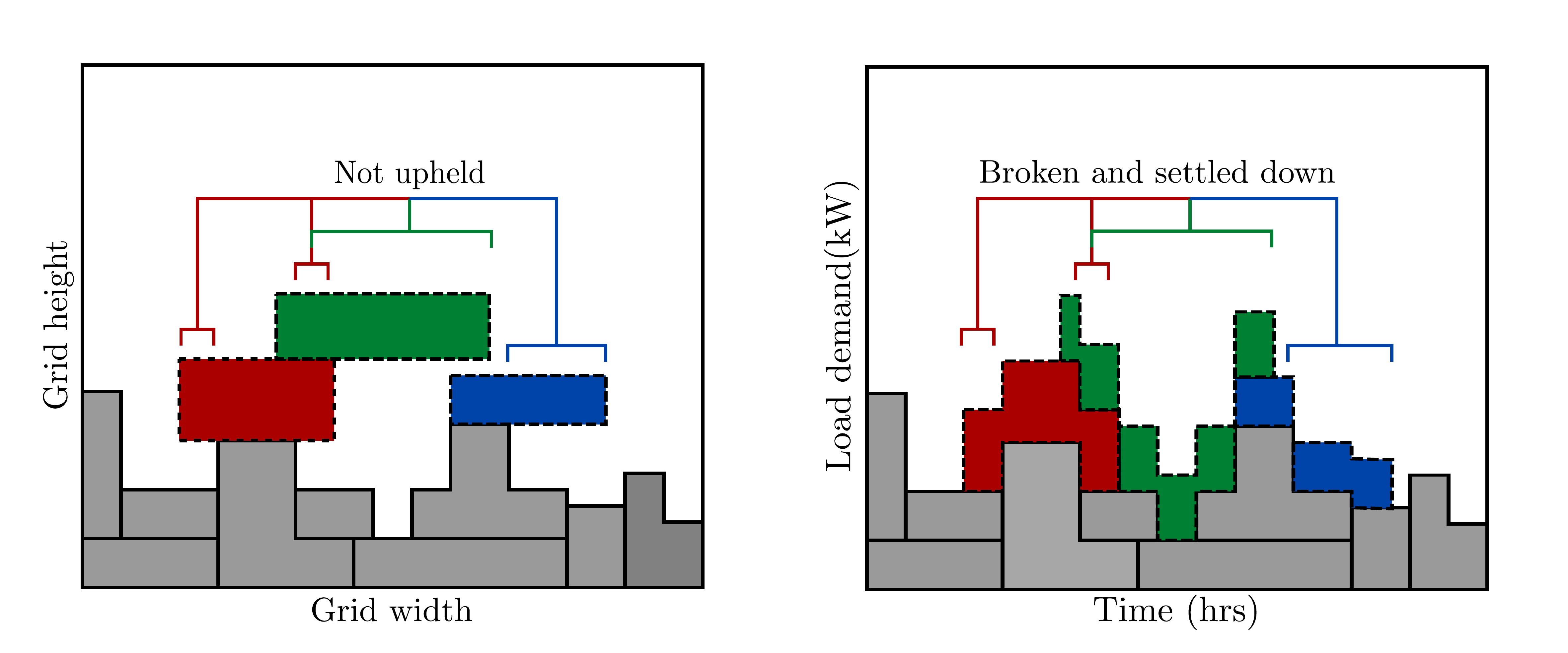}
	\caption{Demonstrates the difference in block settling in game environment (left) and proposed DR simulation (right). Red, Green and Blue blocks aren't supported by the grid base or another load block, thus they are allow to be broken down and slid down to lower level in the simulation. Figure best viewed in color.}
	\label{fig:game_sim}
\end{figure*}

\section{Methods}

Reinforcement learning has shown tremendous strides in learning to take action in game environments surpassing humans. Bringing RL capability to DR created the need to model DR into a game environment. An Atari game called Tetris falls in very close with the simulation environment needed for DR. In Tetris, the player is asked to move blocks of different sizes and shapes in a 2D grid. Actions the player can take on the block are rotation, moving left, right, or down. The player is rewarded with points when a full line in the grid is filled, and the game terminates when the block placed touches the maximum height of the grid. We adapted this game environment to build a simulation to perform DR. The blocks in the game will be device loads in the DR simulation. The player will be replaced by an RL agent who will take action like moving load blocks down, left, and right. Unlike solid blocks in the game environment, the load blocks in the DR simulation are flexible solids, i.e., if part of the load in the grid is not supported by the grid base or another load block, it is allowed to slide down to the lower level as shown in Figure~\ref{fig:game_sim}. The agent reward is determined by the load peak generated when a block is placed in the simulation grid. A positive reward is given if it doesn't increase the current maximum load peak and a negative reward when the current maximum load peak increases. The simulation ends when the load placed crosses the maximum height, which motivates the RL agent to place more load in the simulation grid without generating peaks.

\subsection{Simulation Design}
The simulation is modeled to represent loads on a timescale of 24 hours. The simulation environment consists of 24 columns, each column representing an hour on the 24-hour clock. The height of each column depicts the aggregate load on the electrical system at that particular hour. Ideally the height of the simulation grid is decided by the maximum aggregate load from the historical load profiles of the consumers in a specific grid. As a proof of concept, a maximum of 25kW of the aggregate load is set while training and testing, this can be scaled according to the size of the power grid. If the aggregate load at any hour exceeds the constraints on maximum load, the simulation terminates. 

\subsection{Delineating States and Actions}\label{rl_state}

The learning algorithm stores data in the form of (\textit{state}, \textit{action}, \textit{next state}, \textit{reward}) tuples. The state is an image of the current simulation screen. The action space is defined in terms of three actions: \textit{left}, \textit{right} and \textit{drop} which imply left move, right move and dropping the block onto the simulation respectively. All actions are single actions; i.e., for any state transition, only a single action can be performed, there are no cases where a combination of the individual actions can be performed for a state transition. Taking action at any state causes the game to transition to a new state that generates a reward.

At any point of time if the agent decides on one of the actions the simulation moves to a new state. A right action shifts the current load one cell (representing timestamp) to the right and the new state is the load block shifted one block to the right. A left action shifts the block of load one cell to the left and the simulation transitions to a new state with the load block shifted one cell to the left. A drop action results in the load block being placed on the load profile and a subsequent state transition with the load block now placed on the load profile immediately below the load block. Actions in this sense are discrete and not continuous. At any point of time the state of the simulation is the load profile and the load block. This is captured on every action. The state transitions are finite as there can be finite shapes of blocks and they can be placed in on 24 timestamps. State transitions are deterministic in the the sense that given a state, block, action we can always predict the next state.

\subsection{Deep-Q-Learning (DQN)}\label{dqn}

DQN \cite{c37} agent based on the idea of using a neural network to approximate the below Q as shown in Equation~\ref{eq:1}, and the pipeline of this method can be found in  Algorithm \ref{alg:dqn_algo}.

\begin{equation}\label{eq:1}
\begin{aligned}
Q^{\pi}(s, a) = \max_{\pi} \mathbb{E}_{\pi}[ r_{t} + \gamma r_{t+1} +\gamma ^ {2} r_{t+2} + \\ ..... | s_{t}=s, a_{t}=a]
\end{aligned}
\end{equation}
where $a_{t}$ is the action and $s_{t}$ is the state at time t, $r_{t}$ is the reward obtained by taking action at given state $s_{t}$, $\pi$ denote policy function defines the learning agent's way of behaving at a given time and $\gamma$ is the discount factor. The Q-function can be simplified as:

\begin{equation}\label{eq:2}
Q^{*}(s, a) = \mathbb{E}_{s'} [ r + \gamma \max_{a'} Q^{*}(s', a') | s, a]
\end{equation}
where $s'$ is the state generated by performing action $a$ in state $s$ and $a'$ denoted the action taken in state $s'$.
The DQN model used in the proposed method is a Double DQN \cite{c6}. Double DQNs handles the problem of the overestimation of Q-value. This consists of two networks, namely a policy network and a target network where all changes to the gradients are made on the policy network, which is synced with the target network at regular intervals of episodes. Episode is the length of the simulation at the end of which the system ends in a terminal state. DQN network is used to select the best action to take for the next state (the action with the highest Q-value). The target network is used to calculate the Q-value of taking that action at the next state.

\begin{figure}[!h]
	\centering
	\includegraphics[width=1\columnwidth]{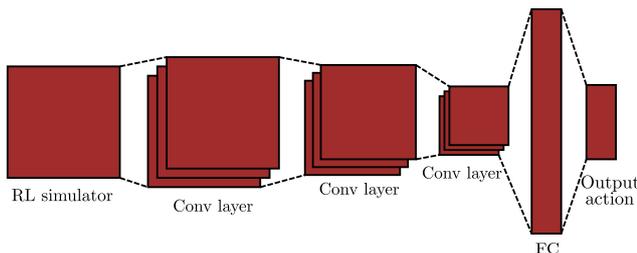}
	\caption{Network architecture used for agent. The policy network and target network have the same architecture. $Convlayer$ consist of convolution block, batch normalization layer and ReLU activation. $FC$ stands for fully connected layer.}
	\label{fig:network}
\end{figure}

The network consists of three convolution layers and one fully connected layer as shown in Figure~\ref{fig:network}. All convolution layers (CNN) use a $5\times5$ kernel and a stride of 2 with batch normalization and ReLU (Rectified Linear Unit) as the activation function. The last hidden layer is a fully connected (FC) layer with 192 units, and the output layer consists of three units for three actions. CNN is used over other network architectures because CNN is memory efficient and best for feature extraction. The number of parameters in FC drastically increases with the increase in hidden layers when compared to CNN, thus increases computation and memory used. Designing a faster model is crucial for applications like load shifting, thus memory efficient and fast neural networks like CNNs are used. According to the design of the proposed RL state, CNN is more apt as it is efficient in feature extraction from raw data with spatial structures like simulation screens. We have also done an extensive analysis of how the selected CNN network over other affects the load shifting problem in the proposed model, as shown in Table~\ref{tab:shallow} and Figure~\ref{fig:analysis_buffer}. Even though the shallow network was able to minimize cost, it creates high load peaks.
With batch normalization, the model normalizes the input layer by adjusting and scaling the activation. Batch normalization reduces the amount by what the hidden unit values of the network shift around (covariance shift), and this is proven in \cite{ioffe2015batch} to speed up the learning process. RMSProp \cite{igel2000improving} is used for optimization in the network.

For training, the network approximates the Q-values for each action and take action with the maximum Q-value. The target network estimate the Q-values of the optimal Q-function Q* by using the Bellman Equation~\ref{eq:2}, since Q-function for any policy obeys the Bellman equation. Temporal difference error $\delta$ is computed by taking the difference of predicted Q-value and the optimal Q-value computed from the Bellman equation using the target network.

\begin{equation}\label{eq:3}
\delta = Q(s, a) - ( r + \gamma \max_{a'} Q^{*}(s', a'))
\end{equation}

By minimizing the loss between the Q-value and the optimal Q-value, the agent arrives at the optimal policy.
Huber loss is used to minimize the error defined as below:

\begin{equation}\label{eq:4}
\mathcal{L} = \frac{1}{|B|} \sum_{b \epsilon B} \mathcal{L}(\delta)
\end{equation}
where batches $B$ of experiences (knowledge acquired in the past) where sampled from the agent's replay buffer and $\mathcal{L}(\delta)$ is defined as 

\begin{equation}\label{eq:5}
\mathcal{L}(\delta)= 
\begin{cases} 
\frac{1}{2}\delta^{2} & |\delta|\leq 1 \\
|\delta|-\frac{1}{2} & Otherwise
\end{cases}
\end{equation}

The Huber loss is robust to outliers. When the error is small, it takes the definition of mean squared error but acts like the mean absolute error when the error is significant.

\begin{figure}
	\input{algo/algo.tex}
\end{figure}

\subsection{Epsilon Greedy Policy and Experience Replay}

Epsilon greedy policy and Experience replay are two crucial techniques that help the learning process of the RL agent drastically. Given the fact that the state space is vast, the agent is enabled to explore the available state space initially without taking too many predicted actions from the network. When the agent takes a random step at any given state its called exploration and when it uses already accumulated experience to make a predicted decision from the network is called exploitation. Exploration and exploitation should not be run exclusively. The agent explores many states at first by taking random actions, but with each successive epoch, it increases the number of informed decisions exploiting the known information to maximize the reward. The decay function $\gamma$ used for this purpose is the exponential decay function which is defined below:

\begin{equation}\label{eq:6}
\gamma = \epsilon_{e} + \frac{\epsilon_{s} +  \epsilon_{e}}{e^{\frac{sd}{\epsilon_{d}}}}
\end{equation}

where $sd$ is the total number of iterations till now and $\epsilon_{d}$ is a hyperparameter controlling the rate of decay from $\epsilon_{s}$ to $\epsilon_{e}$. One iteration translates to taking one action. Note that episodes and iterations have different meanings in this context. Hyperparameters are parameters that are manually set by the practitioner and tuned according to a specific goal.

\begin{table}[ht!]
\centering
    \resizebox{\columnwidth}{!}{
	\begin{tabular}{|l|c|c|c|}
		\hline
		
		Appliances      & Rated power (kWh) & Preferred time & Duration (hrs) \\ \hline
		
		\multicolumn{4}{|c|}{\textbf{Consumer 1}}   \\ \hline
		\multicolumn{4}{|c|}{Non-shiftable}   \\ \hline
		Refrigerator    & 0.5   & 0-24   & 24 \\ \hline
		TV              & 0.5   & 20-23  & 3  \\ \hline
		Indoor lighting & 0.5   & 19-22  & 3  \\ \hline
		Oven            & 1     & 7-8    & 1  \\ \hline
		Stove           & 1     & 12-13  & 1  \\ \hline
		AC              & 1.5   & 13-14  & 1  \\ \hline
		
		\multicolumn{4}{|c|}{Time-shiftable}       \\ \hline
		Washing machine & 1.0, 0.5  & 0-24   & 2   \\ \hline
		Dish washer     & 1.0       & 0-24   & 2   \\ \hline
		Vacuum cleaner  & 1.0       & 0-24   & 1   \\ \hline
		Grinder         & 1.5       & 0-24   & 1   \\ \hline
		
		\multicolumn{4}{|c|}{\textbf{Consumer 2}}   \\ \hline
		\multicolumn{4}{|c|}{Non-shiftable}   \\ \hline
		Refrigerator    & 0.5   & 0-24   & 24 \\ \hline
		TV              & 0.5   & 18-22  & 4  \\ \hline
		Indoor lighting & 0.5   & 18-22  & 4  \\ \hline
		Oven            & 1     & 7-8    & 1  \\ \hline
		Stove           & 1     & 11-12  & 1  \\ \hline
		AC              & 1.5   & 21-23  & 2  \\ \hline
		
		\multicolumn{4}{|c|}{Time-shiftable}       \\ \hline
		Washing machine & 1.0, 0.5  & 0-24   & 2   \\ \hline
		Dish washer     & 1.0       & 0-24   & 1   \\ \hline
		Vacuum cleaner  & 1.0       & 0-24   & 1   \\ \hline
		Grinder         & 1.5       & 0-24   & 1   \\ \hline
		Cloth dryer     & 0.5       & 0-24   & 2   \\ \hline
		
		\multicolumn{4}{|c|}{\textbf{Consumer 3}}   \\ \hline
		\multicolumn{4}{|c|}{Non-shiftable}   \\ \hline
		Refrigerator    & 0.5   & 0-24   & 24 \\ \hline
		TV              & 0.5   & 17-23  & 6  \\ \hline
		Indoor lighting & 0.5   & 18-23  & 5  \\ \hline
		Oven            & 1     & 6-7    & 1  \\ \hline
		Stove           & 1     & 12-14  & 2  \\ \hline
		AC              & 1.5   & 22-24  & 2  \\ \hline
		
		\multicolumn{4}{|c|}{Time-shiftable}       \\ \hline
		Washing machine & 1.0, 0.5  & 0-24   & 2   \\ \hline
		Dish washer     & 1.0       & 0-24   & 2   \\ \hline
		Vacuum cleaner  & 1.0       & 0-24   & 1   \\ \hline
		Cloth dryer     & 0.5       & 0-24   & 1   \\ \hline
		
		\multicolumn{4}{|c|}{\textbf{Consumer 4}}   \\ \hline
		\multicolumn{4}{|c|}{Non-shiftable}   \\ \hline
		Refrigerator    & 0.5   & 0-24   & 24 \\ \hline
		TV              & 0.5   & 18-24  & 6  \\ \hline
		Indoor lighting & 0.5   & 18-23  & 5  \\ \hline
		Stove           & 1     & 13-14  & 1  \\ \hline
		
		\multicolumn{4}{|c|}{Time-shiftable}       \\ \hline
		Washing machine & 1.0, 0.5  & 0-24   & 2   \\ \hline
		Dish washer     & 1.0       & 0-24   & 2   \\ \hline
		Vacuum cleaner  & 1.0       & 0-24   & 1   \\ \hline
		Grinder         & 1.5       & 0-24   & 1   \\ \hline
		Cloth dryer     & 0.5       & 0-24   & 2   \\ \hline
		
		\multicolumn{4}{|c|}{\textbf{Consumer 5}}   \\ \hline
		\multicolumn{4}{|c|}{Non-shiftable}   \\ \hline
		Refrigerator    & 0.5   & 0-24   & 24 \\ \hline
		TV              & 0.5   & 20-24  & 4  \\ \hline
		Indoor lighting & 0.5   & 19-22  & 3  \\ \hline
		Oven            & 1     & 20-21  & 1  \\ \hline
		Stove           & 1     & 12-14  & 2  \\ \hline
		AC              & 1.5   & 12-13  & 1  \\ \hline
		
		\multicolumn{4}{|c|}{Time-shiftable}       \\ \hline
		Washing machine & 1.0, 0.5  & 0-24   & 2   \\ \hline
		Vacuum cleaner  & 1.0       & 0-24   & 1   \\ \hline
		Grinder         & 1.5       & 0-24   & 1   \\ \hline
		Cloth dryer     & 0.5       & 0-24   & 2   \\ \hline
	\end{tabular}
	}
	\caption{Daily load demand of five different consumers. Rated power 1.0, 0.5 indicate  the power rating at first and second hour of the device.}
	\label{tab:load}
\end{table}

\begin{table}[ht!]
	\centering
	\begin{tabular}{|c|c|}
		\hline
		Hyper parameters   & Values \\
		\hline
		$\alpha_1$   & 10.0    \\ \hline
		$\alpha_2$   & 0.76     \\ \hline
		$\alpha_3$   & 0.5    \\ \hline
		$\alpha_4$   & 0.2     \\ \hline
		Buffer size  & $3 \times 10^{4}$     \\ \hline
		Learning rate  & 0.001     \\ \hline
	\end{tabular}
	\caption{Hyper parameters used in experiment.}
	\label{tab:hyperparams}
\end{table}

\begin{table}
	\centering
	\begin{tabular}{|c|c|c|c|}
		\hline
		\multicolumn{4}{|c|}{Price (cents/kWh)}\\
		\hline
		Off-Peak & Mid-Peak & Peak & Off-Peak \\
		\hline
		0-6hrs & 6-15hrs & 15-22hrs & 22-24hrs\\
		\hline
		6 & 9 & 15 & 6 \\
		\hline
	\end{tabular}
	\caption{Monthly energy billing scheme.}
	\label{cost_table}
\end{table}

\begin{figure*}[ht]
	\centering
	\begin{subfigure}{.45\textwidth}
		\centering
		\includegraphics[width=1\linewidth]{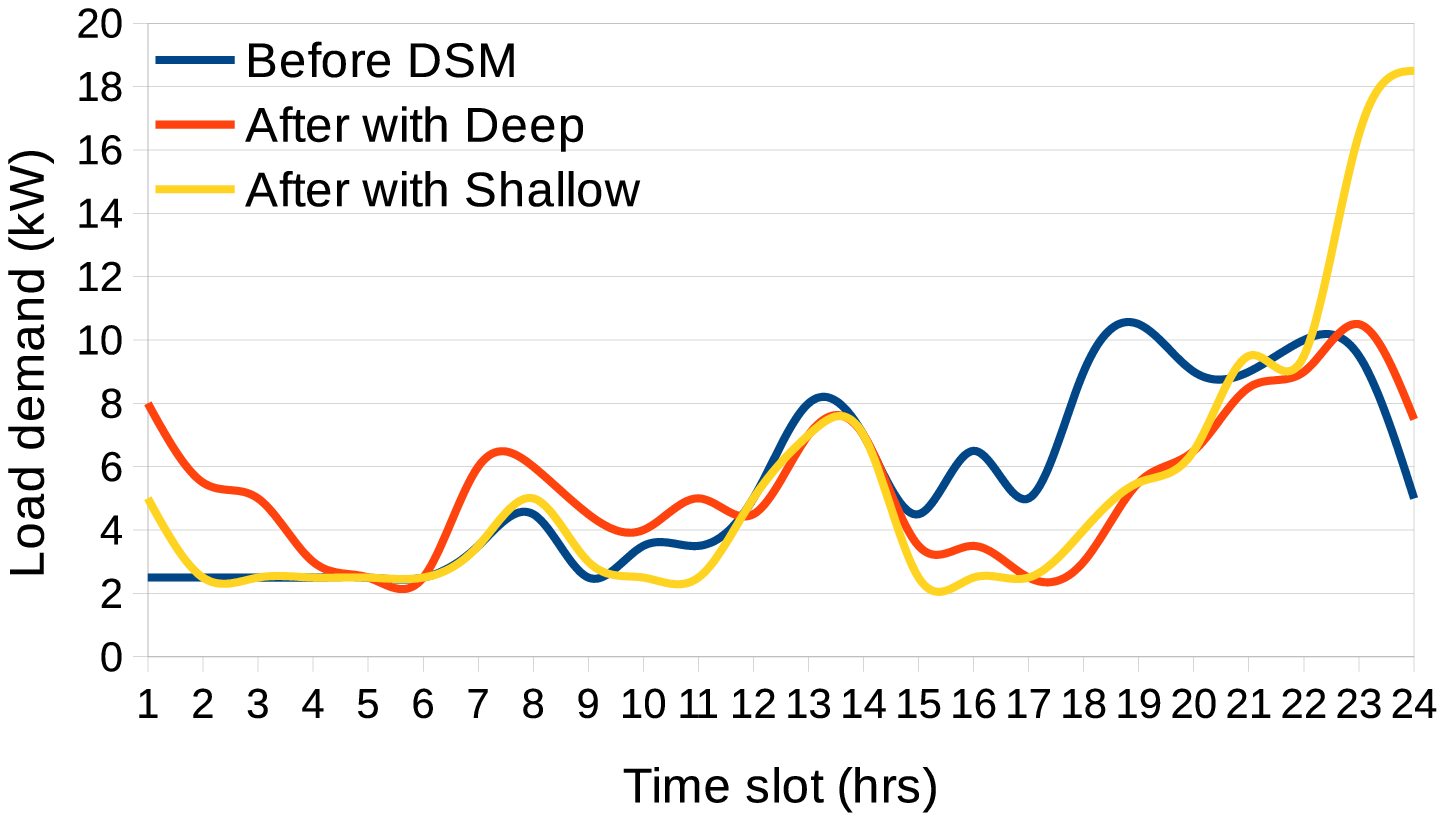}
		\caption{}
		\label{fig:analysis_net}
	\end{subfigure}
	\begin{subfigure}{.45\textwidth}
		\centering
		\includegraphics[width=1\linewidth]{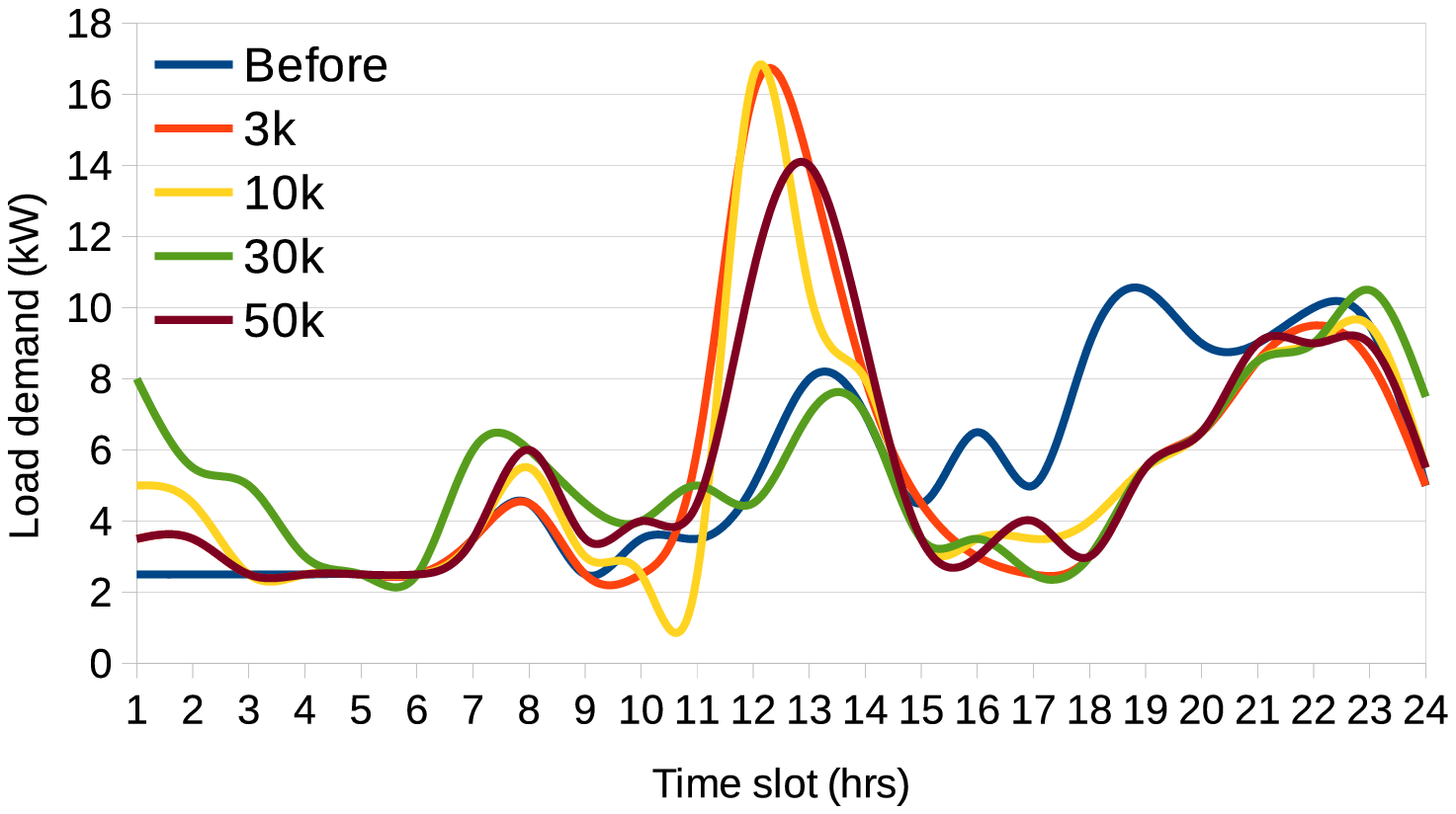}
		\caption{}
		\label{fig:analysis_buffer}
	\end{subfigure} 
	\caption{RL-DSM results\footnotemark[1] of aggregated residential loads of five different consumer to minimize daily peak load and cost (a) with Shallow (After with Shallow) and Deep network (After with Deep) (b) using different memory buffer size like 3000 (3k), 10000 (10k), 30000 (20k) and 50000 (50k). The figure best viewed in color.}
	\label{fig:new_analysis}
\end{figure*}

The Experience Replay \cite{c7} is used to avoid the agent from forgetting previous experiences as it trains on newer states and to reduce correlations between experiences. In Experience Replay, a buffer of the old experiences is maintained and the agent is trained on it. By sampling batches of experience from the buffer at random, the correlation between the experiences can be reduced. The reason for this is that if you were to sample transitions from the online RL agent as they are being experienced, there would be a strong temporal/chronological relationship between them. During our experiments, we fixed buffer size at $3 \times 10^{4}$. We empirically found the optimal buffer size that works best for our model without allotting huge amount of reserved memory space. We have done extensive analysis on picking the right buffer size, which can be found in Table~\ref{tab:buffersize} and Figure~\ref{fig:analysis_buffer}.

\subsection{Reward Systems}\label{reward_peak}

The proposed reward system consists of three important aspects, maximum height $h^{max}$, variance of the height distribution $var(h)$ and number of complete lines $l$ formed from the distribution. The reward system $R$ is summarized as:

\begin{equation}\label{eq:7}
R =  \alpha_1 * var(h) + \alpha_2 * l - \alpha_3 * h^{max}
\end{equation}

The variance of the height distribution $var(h)$ is given by:

\begin{equation}\label{eq:8}
var(h) = \frac{1}{1 + var(h_a)}
\end{equation}

where $var(h_a)$ the variance of the height distribution after taking the action.

The variance of load distribution reward $var(h)$ would encourage the agent to shift load such that more uniform distribution of height of the load profile can be attained. Complete lines $l$ reward complements the agent to increase the spread to have more complete lines (rows) in the simulation. This reward component enables the agent to decide to shift a load to either side edges of the simulation. Maximum high reward component $h^{max}$ helps the agent to avoid making a peak in the load profile as this reward contributes negative reward if the current maximum high increased. Other hyperparameters used for the experiments are shown in Table~\ref{tab:hyperparams}.

The variance of load distribution encountered some cases where the agent realised that placing the blocks on the same position also lead to increase in variance for a sufficiently well spread out distribution. Due to this, initially the agent would perform as expected by placing blocks in a manner which caused the distribution to spread out, but after the certain point it would start placing all the blocks at the same position, resulting in the game terminating with poor results. To counter this, standard deviation was utilized to properly scale the impact of distribution on the overall reward vis-à-vis the other reward parameters.

For simultaneous cost and peak minimization, an additional reward term $c$ is introduced to the peak minimization reward system. The cost is calculated with the help of the piece-wise pricing adapted from \cite{c9} as shown in Table~\ref{cost_table}. As the price of the load schedule increases, the reward for the agent decreases; the term helps the agent to shift load that results in lower billing price, as shown in the Equation~\ref{eq:9}.

\begin{equation}\label{eq:9}
R = \alpha_1 * var(h) + \alpha_2 * l - \alpha_3 * h^{max} - \alpha_4 * c
\end{equation}

\footnotetext[1]{Continuous curves have been used for better visualization instead of the discretized plot even though the load demand is only captured at 24 timestamps.}

\begin{figure*}[ht!]
	\centering
	\begin{subfigure}{.5\textwidth}
		\centering
		\includegraphics[width=1\linewidth]{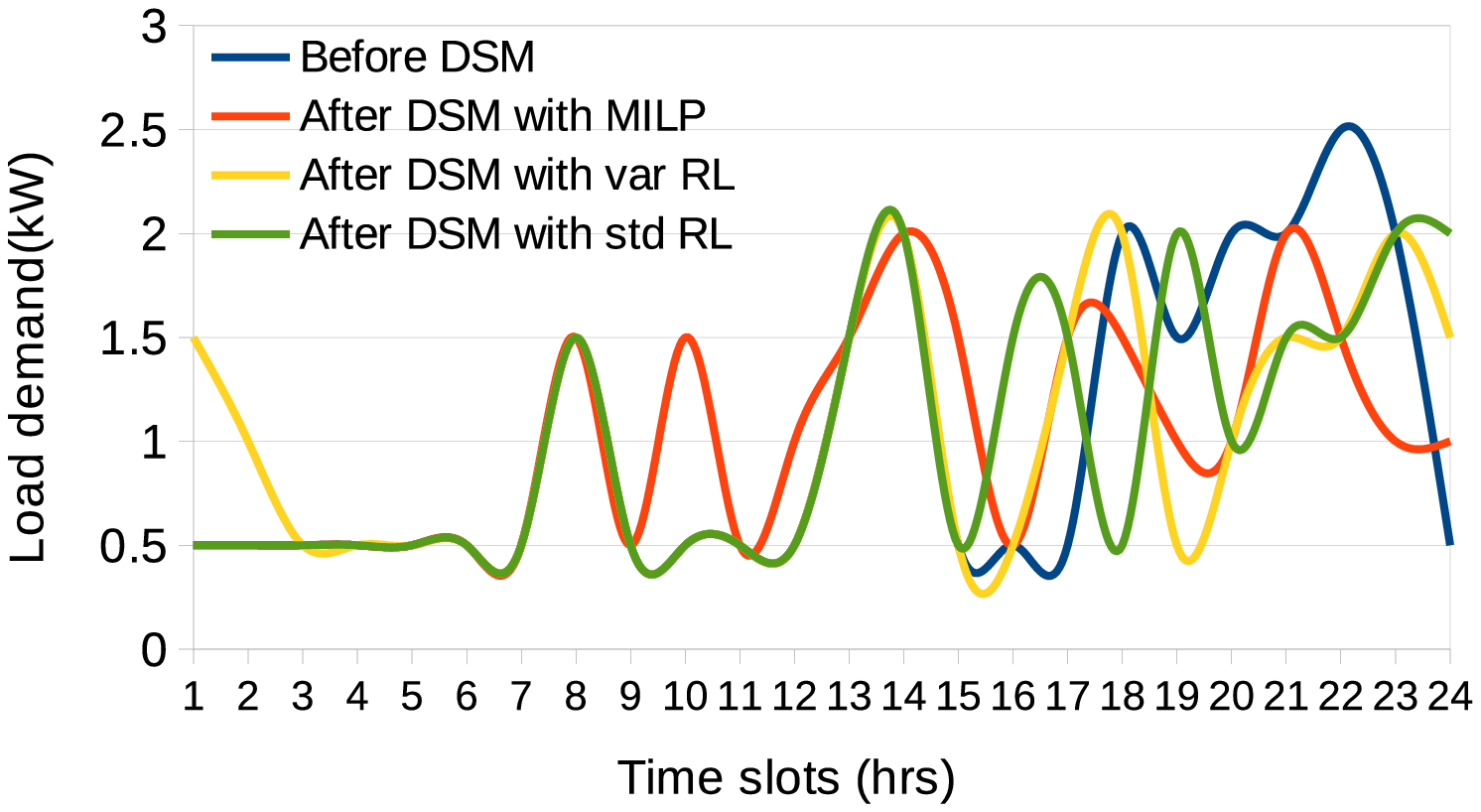}
		\caption{}
	\end{subfigure}%
	\begin{subfigure}{.5\textwidth}
		\centering
		\includegraphics[width=1\linewidth]{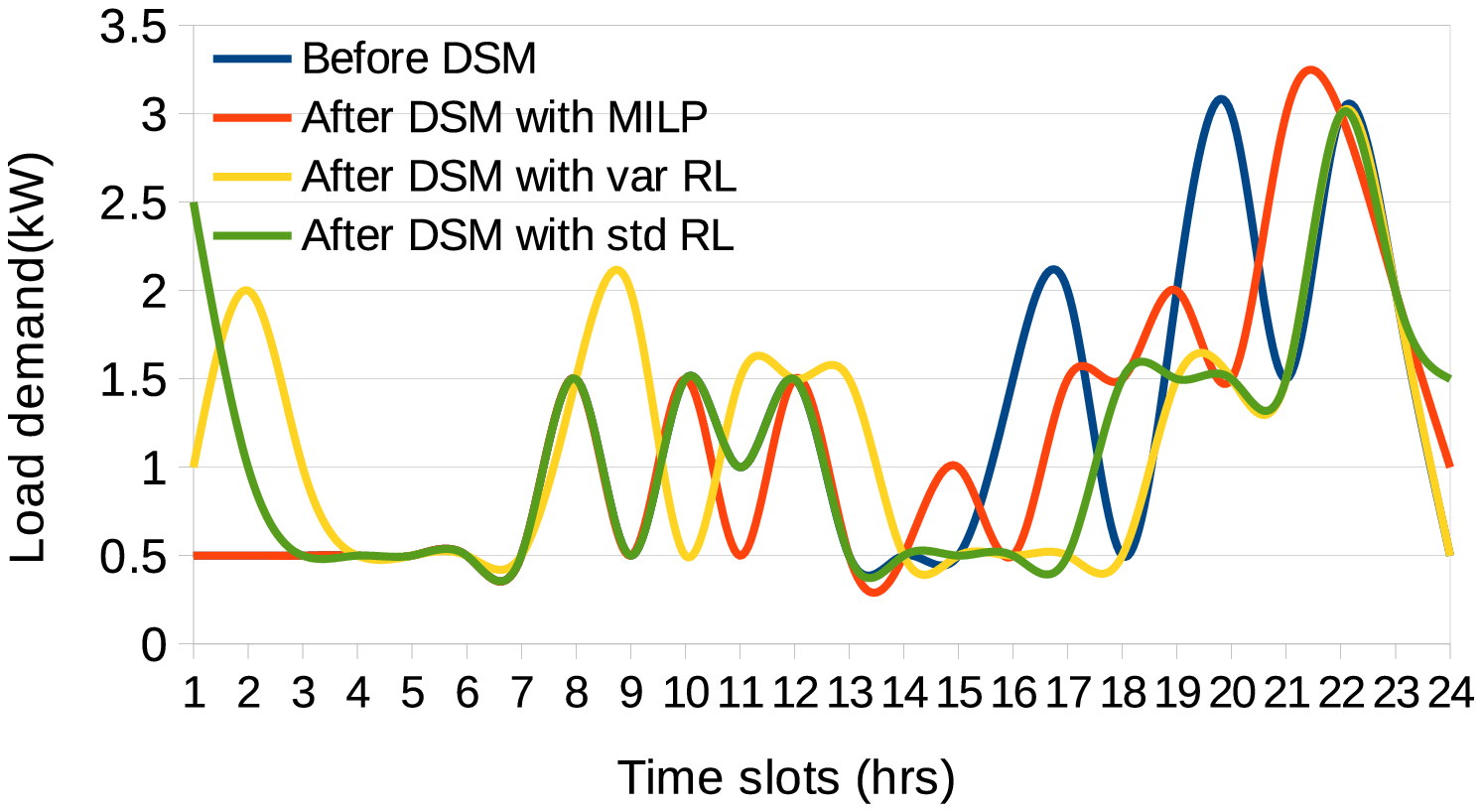}
		\caption{}
	\end{subfigure}
	\begin{subfigure}{.5\textwidth}
		\centering
		\includegraphics[width=1\linewidth]{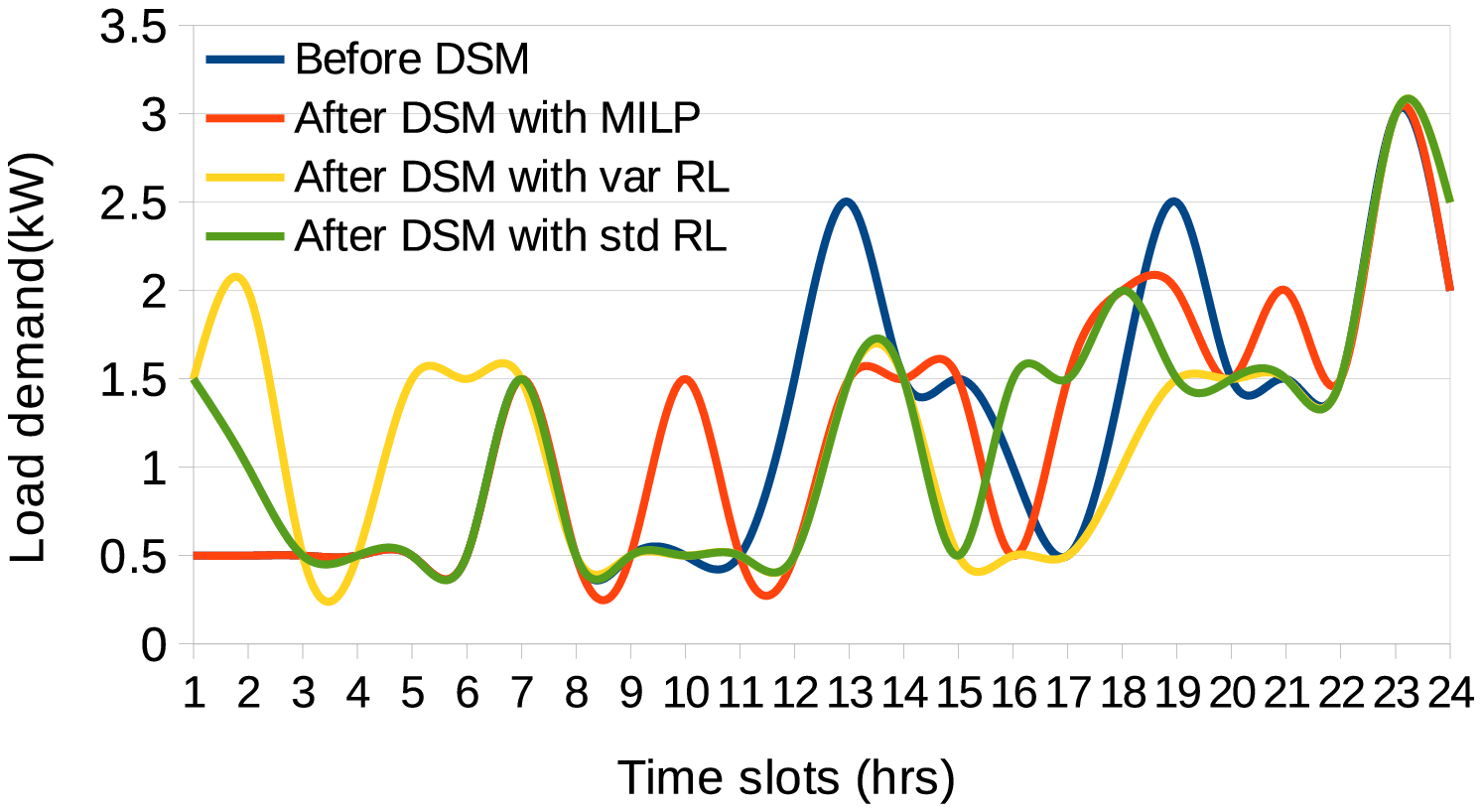}
		\caption{}
	\end{subfigure}%
	\begin{subfigure}{.5\textwidth}
		\centering
		\includegraphics[width=1\linewidth]{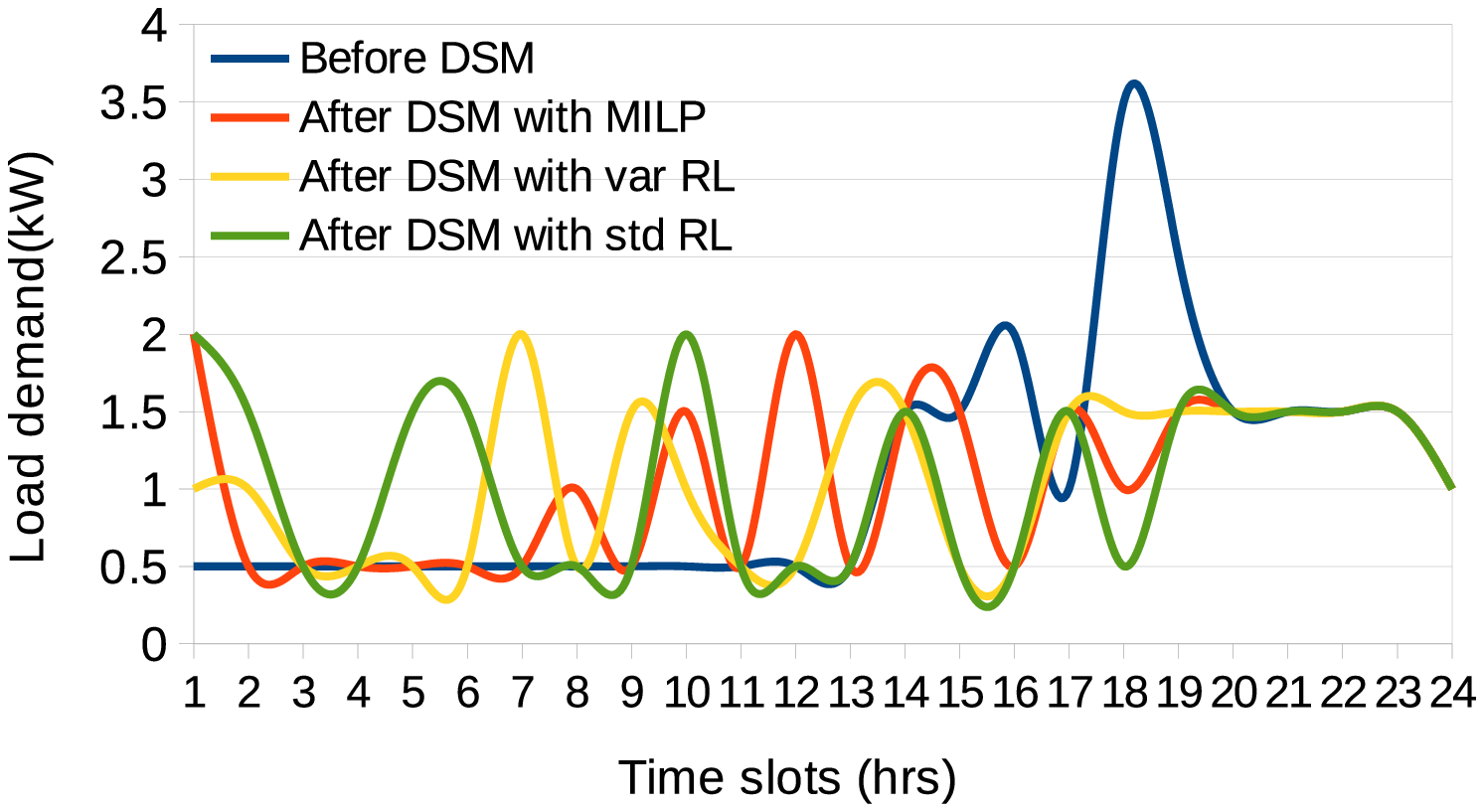}
		\caption{}
	\end{subfigure}
	\begin{subfigure}{.5\textwidth}
		\centering
		\includegraphics[width=1\linewidth]{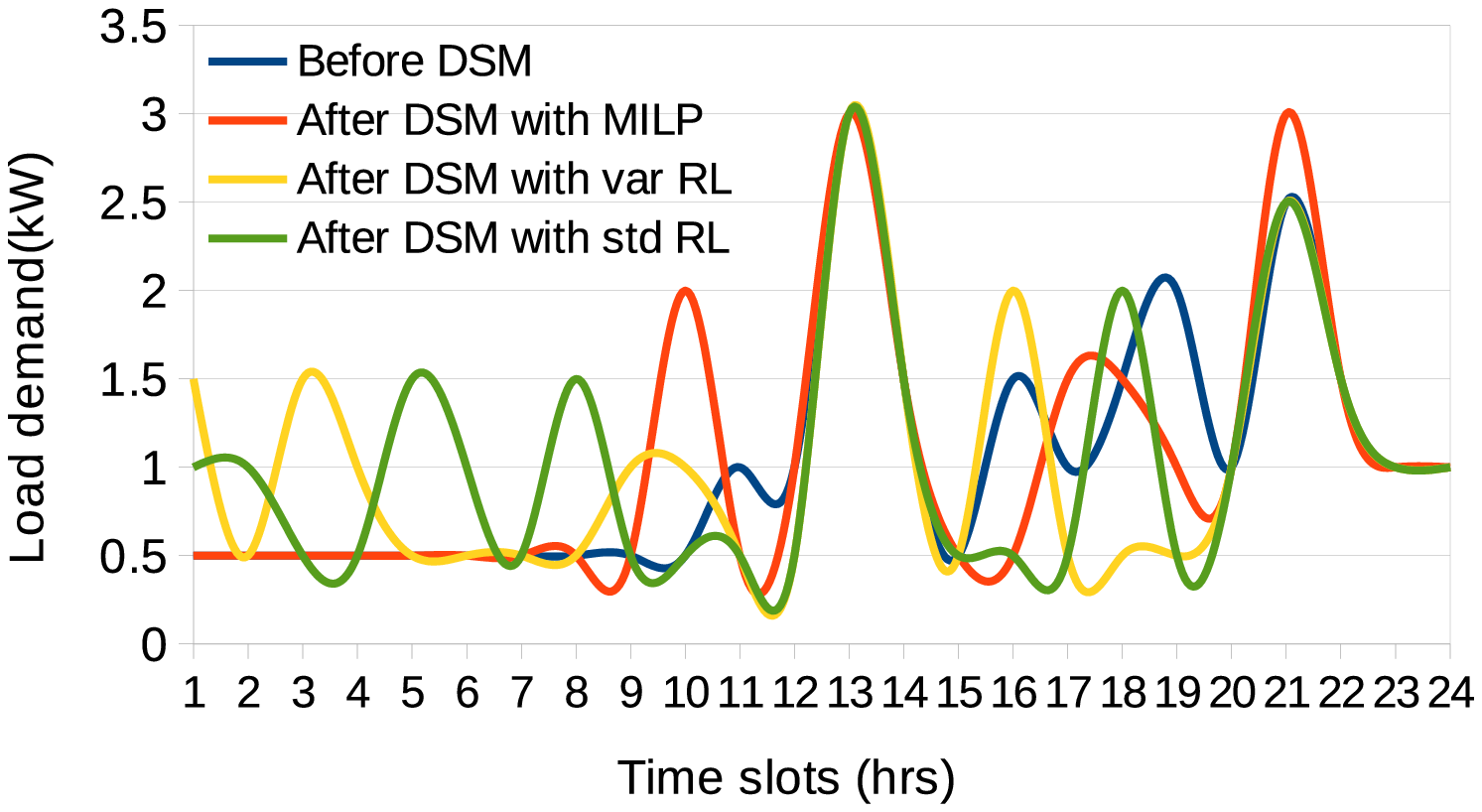}
		\caption{}
	\end{subfigure}
	\begin{subfigure}{.49\textwidth}
		\centering
		\includegraphics[width=1\linewidth]{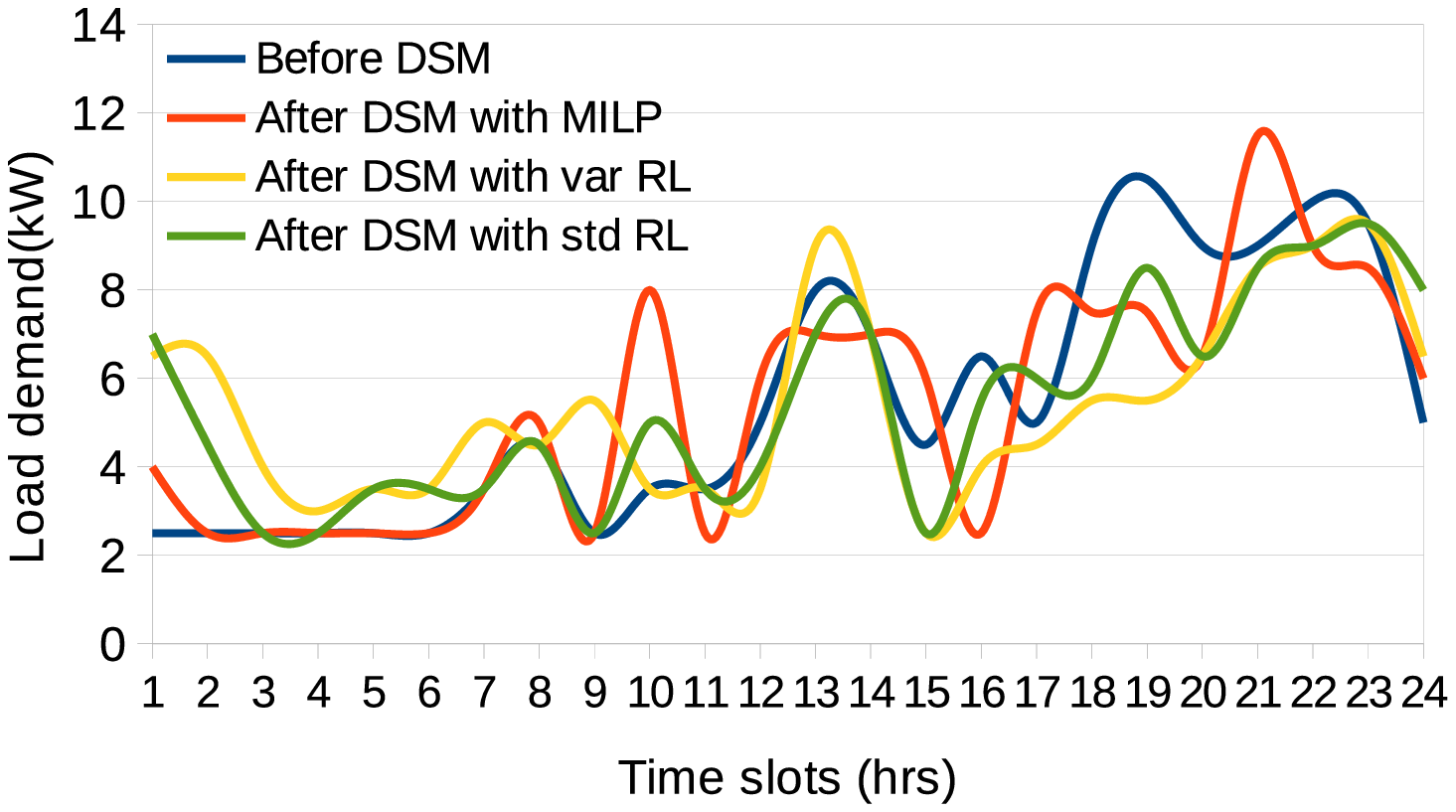}
		\caption{}
	\end{subfigure}
	\caption{RL-DSM results\footnotemark[1] of residential loads for five different consumers (a), (b), (c), (d) and (e) respectively with Case 1 objective of peak minimization which are compared with MILP. (f) is the RL-DSM results of aggregated residential loads for five different consumers to minimize daily peak load with Standard deviation(std RL) and Variance(var RL). The figure best viewed in color.}
	\label{fig:peakmin}
\end{figure*}

\begin{table*}[ht!]
	\centering
	\begin{tabular}{|c|c|c|c|c|c|c|c|}
		\hline
		\multirow{2}{*}{Consumer}    & \multicolumn{4}{c|}{Total energy bill (\$/month)} & \multicolumn{3}{c|}{Monthly savings (\$)} \\ \cline{2-8}
		& without& MILP & RL with std & RL with var & MILP & RL with std & RL with var \\ \hline
		Consumer 1 & 81 	& 77.85 & 76.95  & 74.25  & 3.15  & 4.05    & \textbf{6.75} \\ \hline
		Consumer 2 & 92.25  & 90.5  & 82.8   & 81.9   & 1.35  & 9.45    & \textbf{10.35} \\ \hline
		Consumer 3 & 87.75  & 89.55 & 87.75  & 79.65  & -1.8  & 0       & \textbf{8.1}  \\ \hline
		Consumer 4 & 88.2  	& 78.75 & 80.1   & 80.1   & 9.45  & \textbf{8.1}     & \textbf{8.1}  \\ \hline
		Consumer 5 & 82.8   & 81  	& 82.8   & 78.05  & 1.8   & 0       & \textbf{6.75} \\ \hline
		All        & 432   	& 418.05& 410.4  & 391.95 & 13.95 & 21.6    & \textbf{40.05} \\ \hline
	\end{tabular}
	\caption{Energy bill minimization with peak reduction (Case 1).}
	\label{tab:energypeakmin}
\end{table*}

\begin{figure*}[ht!]
	\centering
	\begin{subfigure}{.5\textwidth}
		\centering
		\includegraphics[width=1\linewidth]{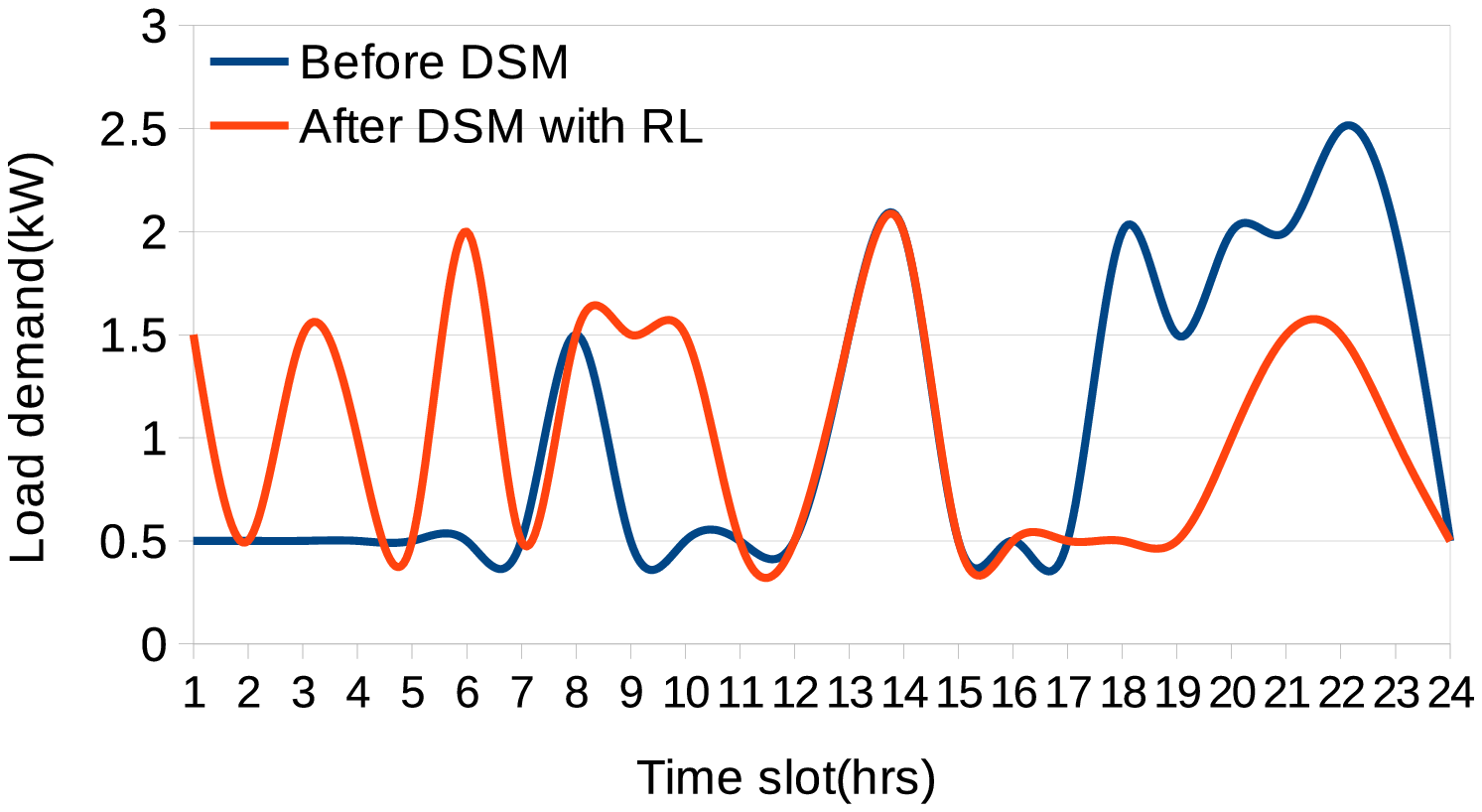}
		\caption{}
	\end{subfigure}%
	\begin{subfigure}{.5\textwidth}
		\centering
		\includegraphics[width=1\linewidth]{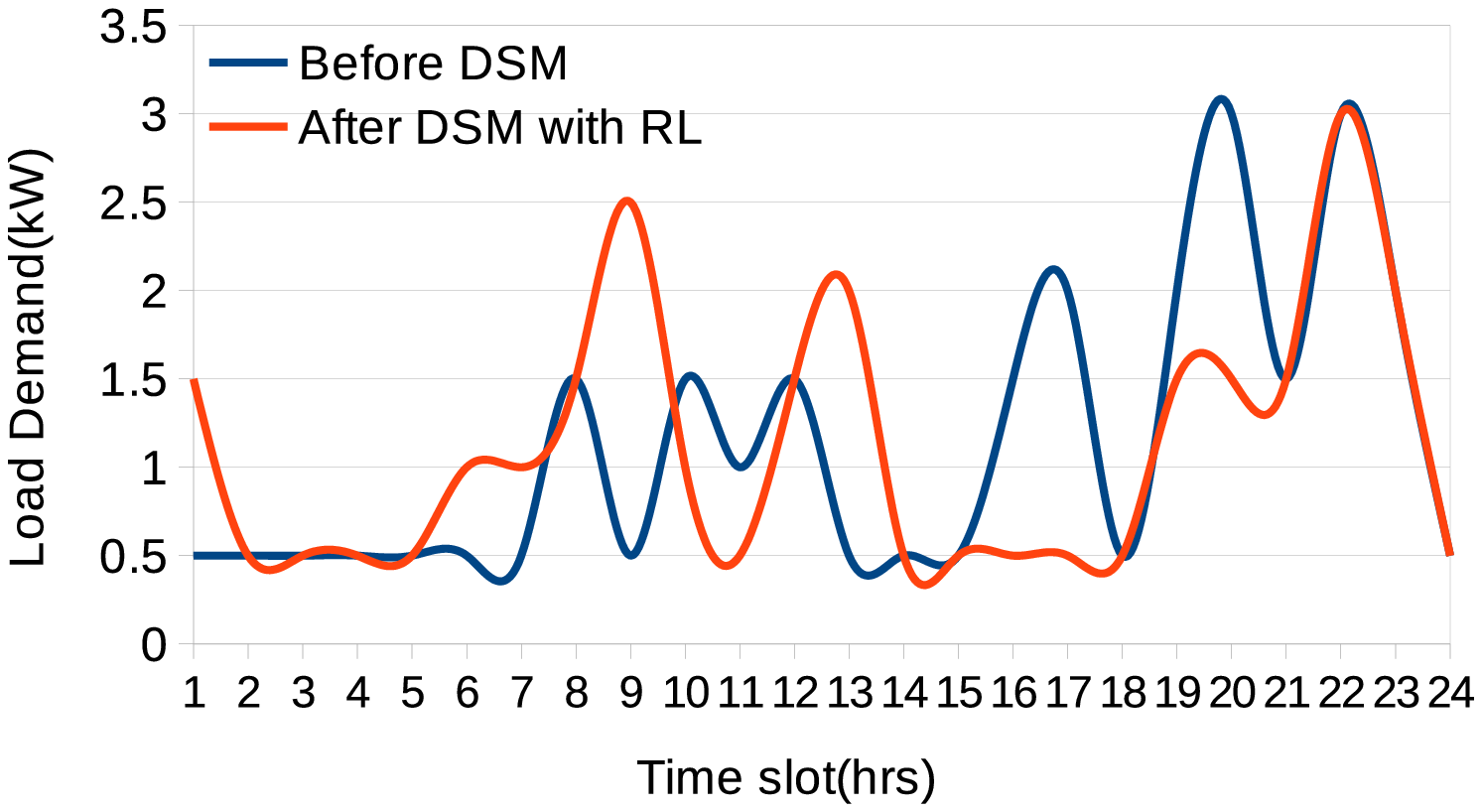}
		\caption{}
	\end{subfigure}
	\begin{subfigure}{.5\textwidth}
		\centering
		\includegraphics[width=1\linewidth]{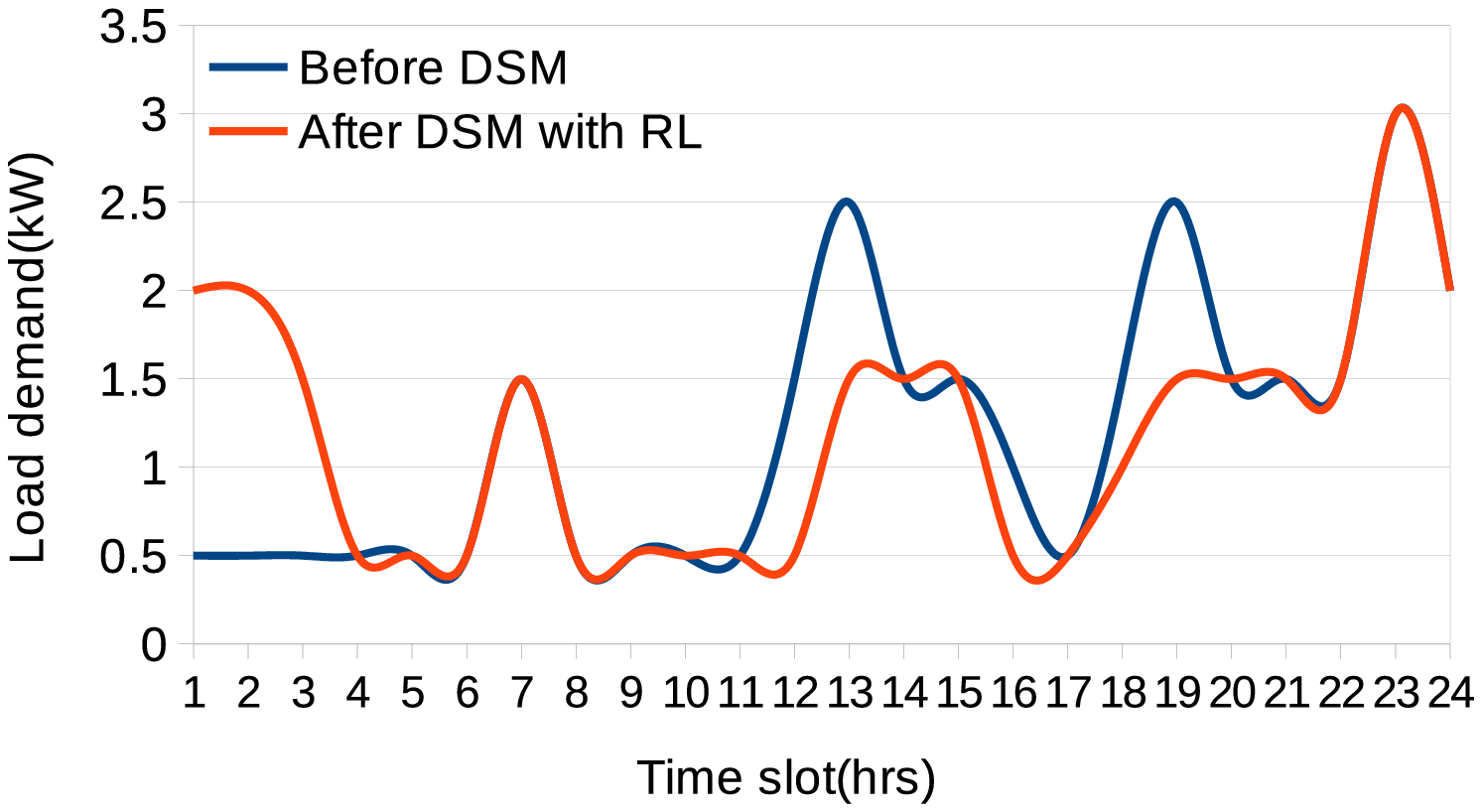}
		\caption{}
	\end{subfigure}%
	\begin{subfigure}{.5\textwidth}
		\centering
		\includegraphics[width=1\linewidth]{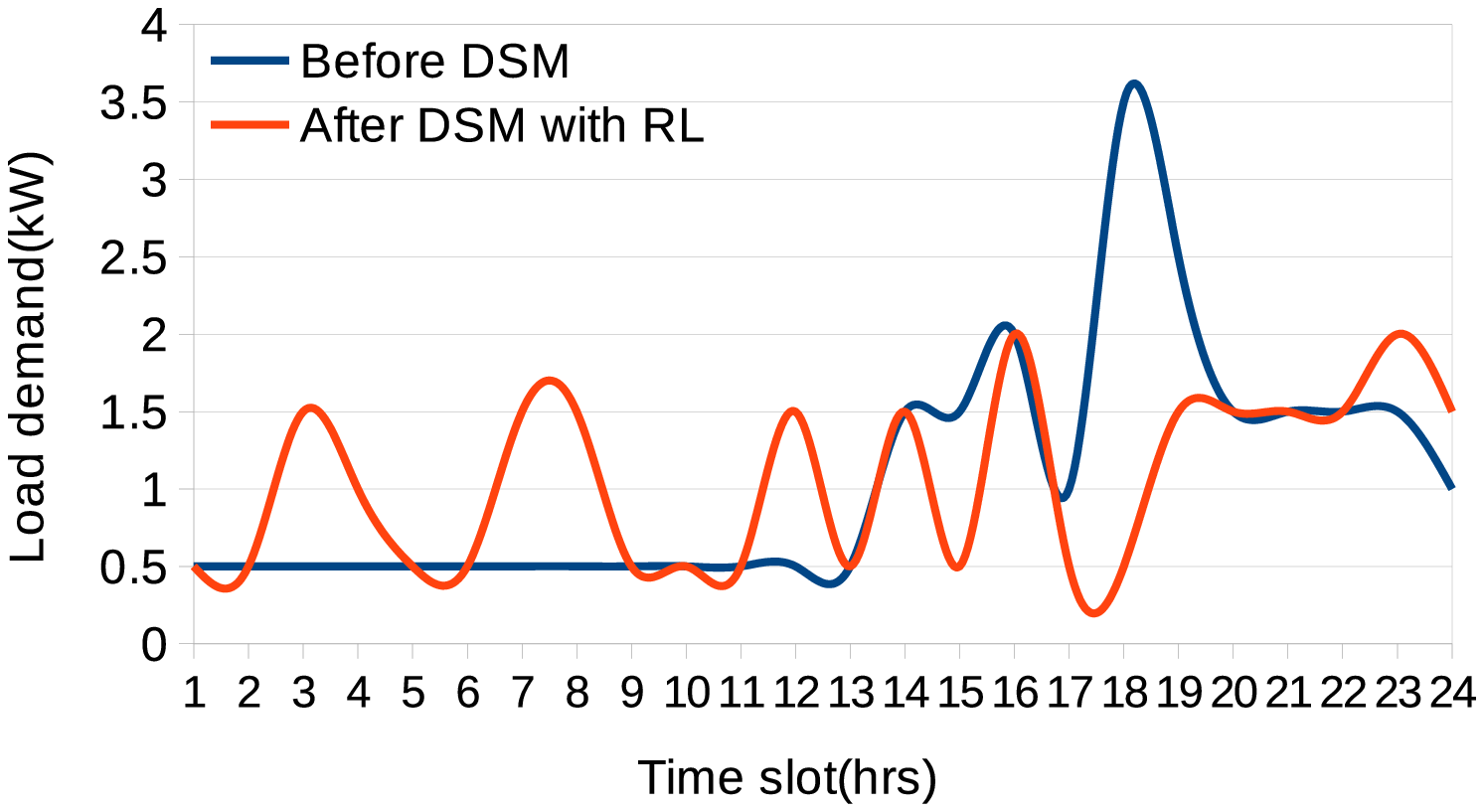}
		\caption{}
	\end{subfigure}
	\begin{subfigure}{.5\textwidth}
		\centering
		\includegraphics[width=1\linewidth]{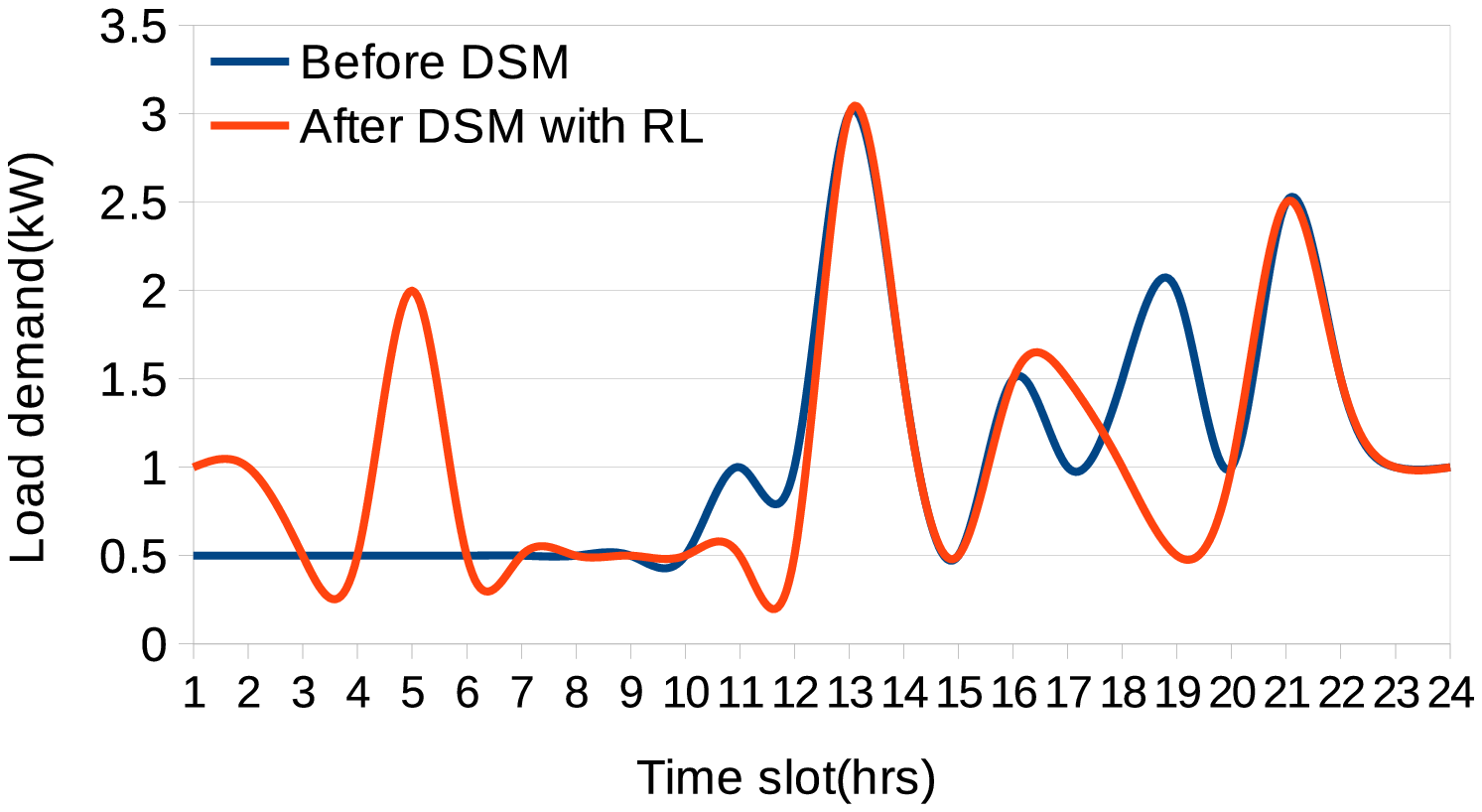}
		\caption{}
	\end{subfigure}
	\begin{subfigure}{.49\textwidth}
		\centering
		\includegraphics[width=1\linewidth]{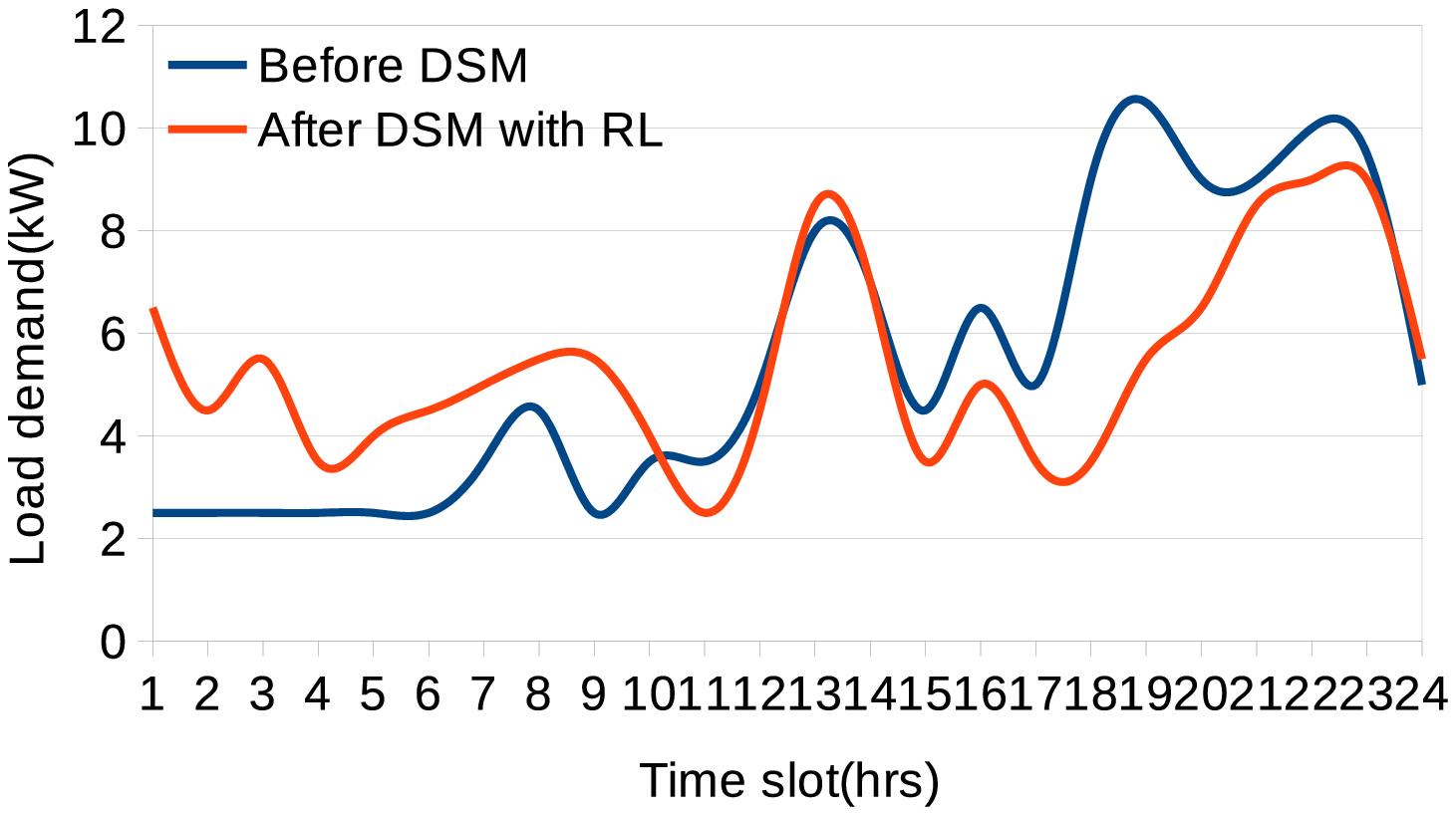}
		\caption{}
	\end{subfigure}
	\caption{RL-DSM results\footnotemark[1] of residential loads for five different consumers (a), (b), (c), (d) and (e) respectively with Case 2 objective of peak and cost minimization. (f) is the RL-DSM results of aggregated residential loads for five different consumers to minimize daily peak load and cost. The figure best viewed in color.}
	\label{fig:peakcostmin}
\end{figure*}

\begin{table*}[ht!]
	\centering
	\begin{tabular}{|c|c|c|c|}
		\hline
		\multirow{2}{*}{Consumer}   & \multicolumn{2}{c|}{Total energy bill (\$/month)} & \multirow{2}{*}{Monthly savings (\$)} \\
		\cline{2-3}
		& before DSM  & after DSM   &          \\ \hline
		Consumer 1 & 81       & 73.35    & 7.65     \\ \hline
		Consumer 2 & 92.25    & 81       & 11.25    \\ \hline
		Consumer 3 & 87.75    & 79.65    & 8.1      \\ \hline
		Consumer 4 & 88.2     & 75.15    & 13.05    \\ \hline
		Consumer 5 & 82.8     & 72.45    & 10.35    \\ \hline
		All        & 432      & 381.6    & 50.4     \\ \hline
	\end{tabular}
	\caption{Energy bill minimization with peak and cost reduction (Case 2).}
	\label{tab:energypeakcostmin}
\end{table*}

\begin{table*}[ht]
	\centering
	\begin{tabular}{|c|c|c|c|c|c|c|c|c|}
		\hline
		\multirow{2}{*}{Consumer}   & \multicolumn{3}{c|}{Total energy bill (\$/month)} & \multicolumn{2}{c|}{Monthly savings (\$)}  & \multicolumn{2}{c|}{Peak load (kW)} \\ \cline{2-8}
		           & Before  & Shallow Net   & Deep Net  & Shallow Net & Deep Net & Shallow Net & Deep Net    \\ \hline
		Consumer 1 & 81 	  & 71.55     & 73.35  & 9.45    & 8.1   & 4              & \textbf{2} \\ \hline
		Consumer 2 & 92.25    & 82.35     & 81     & 9.9     & 13.5  & \textbf{3}     & \textbf{3} \\ \hline
		Consumer 3 & 87.75    & 79.65     & 79.65  & 8.1     & 3.6   & 5              & \textbf{3} \\ \hline
		Consumer 4 & 88.2  	  & 71.1      & 75.15  & 17.1    & 14.4  & 5.5            & \textbf{2.5} \\ \hline
		Consumer 5 & 82.8     & 72.45  	  & 72.45  & 10.35   & 3.6   & \textbf{3}     & \textbf{3} \\ \hline
		All        & 432   	  & 377.1     & 381.6  & 54.9    & 43.2  & 18.5           & \textbf{10.5} \\ \hline
	\end{tabular}
	\caption{Demonstration of deep CNN network's efficiency compared to shallow network. Shallow Net: Shallow network and Deep Net: Deep network.}
	\label{tab:shallow}
\end{table*}

\begin{table*}[ht]
	\centering
	\begin{tabular}{|c|c|c|c|c|c|c|c|c|}
		\hline
		\multirow{2}{*}{Consumer}   &\multicolumn{4}{c|}{Monthly savings (\$)}  & \multicolumn{4}{c|}{Peak load (kW)} \\ 
		\cline{2-9}
		           & 3k      & 10k             & 30k              & 50k    & 3k         & 10k    & 30k           & 50k \\ \hline
		Consumer 1 & 8.1     & \textbf{9.45}   & 7.65             & 8.1    & 4          & 3      & \textbf{2}    & 4.5 \\ \hline
		Consumer 2 & 7.2     & 7.2             & \textbf{11.25}   & 8.1    & 4.5        & 4.5    & \textbf{3}    & \textbf{3} \\ \hline
		Consumer 3 & 2.7     & 0.45            & \textbf{8.1}     & 2.7    & \textbf{3} & 4      & \textbf{3}    & \textbf{3} \\ \hline
		Consumer 4 & 12.6    & \textbf{13.5}   & 13.05            & 9      & 3.5        & 3      & \textbf{2.5}  & \textbf{2.5} \\ \hline
		Consumer 5 & 7.2     & 7.2             & \textbf{10.35}   & 8.1    & 4          & 4.5    & 3             & \textbf{2} \\ \hline
		All        & 37.8    & 37.8            & \textbf{50.4}    & 36     & 16         & 16.5   & \textbf{10.5} & 14 \\ \hline
	\end{tabular}
	\caption{Demonstration of the effect of different memory buffer size like 3000 (3k), 10000 (10k), 30000 (20k) and 50000 (50k) in RL-DSM for peak and cost minimization.}
	\label{tab:buffersize}
\end{table*}

\section{Simulation Results and Discussion}

The proposed RL model is tested in two different case studies with five customers load data adapted from \cite{c12}, as shown in Table~\ref{tab:load}.  In case 1, the utility energy bill is reduced with peak minimization objective. Case 2 reduces the utility energy bill by lowering peak and cost simultaneously. The models based on the proposed peak minimization technique showcase better results than MILP. 

\subsection{Case 1}

The objective of this case study is to minimize peak load demand with RL and thereby reducing the energy bill of the consumer. The model is tested with one objective on five different consumers, as discussed in the above sections. DR with RL (using DQN agent) scheduled load is compared with MILP scheduling for all consumers in 
Figure~\ref{fig:peakmin}. Figure~\ref{fig:peakmin} shows how the system reacts to the aggregated load of all consumers. The effect of selecting the right reward function of the RL model for scheduling each consumer load is shown in Table~\ref{tab:energypeakmin} by comparing variance and standard deviation as a reward function. The variance reward system outperformed the standard deviation for all test consumer data.

As shown in Table~\ref{tab:energypeakmin}, the aggregated load reduced the overall bill cost from \$432 to \$391.95, saving around 9.27\%. From the results, it can be inferred that the proposed method could reduce the monthly bill dramatically when compared to other traditional methods on single peak objective minimization.

\subsection{Case 2}

In this case study, two different objectives are minimized simultaneously with RL. Here, the model reduces the peak load and cost of the load schedule. This hybrid multi-objective function guarantee that the load profile does not have high peaks, and at the same time, the aggregated price of the load per day is minimum, as seen in Figure~\ref{fig:peakcostmin}. It is observed that the loads from peak hours have been moved to min peak or low peak time to minimize high power demand. Taking load cost into account trains the model to understand that moving most loads from peak to non-peak hours is not entirely ideal for the consumer as it may not shift the load to the time duration at lower prices. Adding the cost factor helps the agent to move the blocks to the lower-priced time duration, which currently has fewer loads. However, the experiments with single cost objective, which only considers cost and disregards peak minimization entirely, shows that non-peak hours started to get high peak loads as expected because the reward system for the agent focused solely on cost minimization. This multi-objective function solves this issue. Adding peak minimization with cost minimization assures that the agent is not fixated on cost minimization, causing peaks to be formed during a time when the cost of electricity consumption is less. Thus, both the parameters of the multi-objective function, peak and cost minimization are mutually beneficial for one another. Figure~\ref{fig:peakcostmin} shows how the system reacts to the aggregated load of all consumers.

As shown in Table~\ref{tab:energypeakcostmin}, the aggregate load reduced bill cost from \$432 to \$381.6, saving around 11.66\%. The current limitation of the proposed system is that it does not support dynamic preferred time for each device, and load demand is on the scale of 0.5kW. This is the reason that the daily test load demand shown in Table~\ref{tab:load} has a 1hour/unit time scale and 0.5kW/unit power demand scale to fit the proposed RL simulator.

\begin{figure}
	\centering
	\includegraphics[width=1\columnwidth]{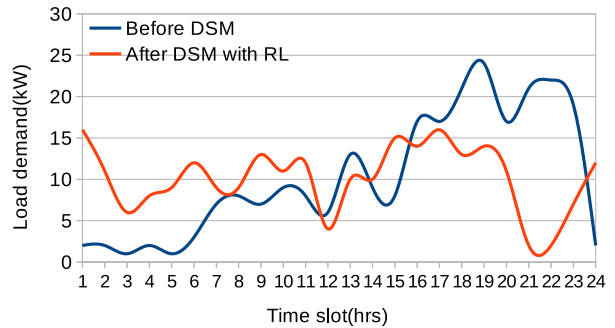}
	\caption{RL-DSM results\footnotemark[1] on loads of 14 different devices to demonstrate the scalability of the proposed method.}
	\label{fig:heavypeak}
\end{figure}

As traditional methods like MILP does not have a holistic knowledge of the system, it can be only used for a fixed number of devices. If the traditional system needs to accommodate more devices at some point in time after deployment, the method needs to be rerun, and scheduling factors should be computed again. The proposed methods are more flexible as it has the learning capability to handle any load given to it. Even after the proposed RL model is deployed, there is no need for any adjustment to add or remove devices from the system at any time instant. The time complexity of the proposed method is better than other traditional methods and its linearly proportional to number of devices, unlike traditional methods that have exponential growth in time concerning the number of devices to be scheduled. A trained RL agent always operates with a fixed number of floating-point arithmetic (FLOPs) and multiply-adds (MAdd) for any number of loads to be scheduled. The number of FLOPs and MAdd only depend on the architecture of the network. This makes the proposed model schedule hundreds of devices time and space-efficient than traditional methods. The time complexity of RL can be defined as $\mathcal{O}(kp)$ and MILP would be $\mathcal{O}(k^2p)$ where $p$ is the number of parameters associated with the model and $k$ is the number of devices to be scheduled. The space complexity of RL can be formulated as $\mathcal{O}(p)$, whereas for MILP would be $\mathcal{O}(k^2)$. The scalability of the proposed methodology is demonstrated by scheduling 46 loads of 14 different devices but with power-hungry devices with RL agent as shown in Figure \ref{fig:heavypeak}. Deploying trained RL agents in an embedded device is as easy as deploying traditional models.

\section{Conclusion}

The exponential growth in demand for power in the household has increased the stress on the power grid to meet its demand. DR can help a smart grid to improve its efficiency to meet the power need of the customer. This paper proposes a residential DR using a deep reinforcement learning method where both load profile deviation and utility energy bills are minimized simultaneously. An extensive case study with a single objective and multi-objective cases is conducted. In both cases, the proposed method outperformed the traditional MILP method. This paper exhibited the potential of reinforcement learning for better smart grid operations and tested all these cases on five different consumers and showcased the ability of the proposed method.  In the future, this work can be extended to introduce variable preferred time for each device and improve the RL simulation grid to accommodate devices with lower power demand with a scale of 0.1kW. This work can also be extended to schedule more granular timed devices with less than 1hr to complete its task. In this work, renewable energy sources and energy storage are not considered. Designing an RL agent to manage renewable energy and energy storage can bring out the full potential of AI models in energy management.
	
\bibliography{refs}


\clearpage
\onecolumn
\begin{center}
	{\Large Supplementary Material:\\ 
		Intelligent Residential Energy Management System using Deep Reinforcement Learning}
\end{center}

\setcounter{section}{0}

\begin{figure*}[!ht]
	\centering
	\begin{subfigure}{.5\textwidth}
		\centering
		\includegraphics[width=1\linewidth]{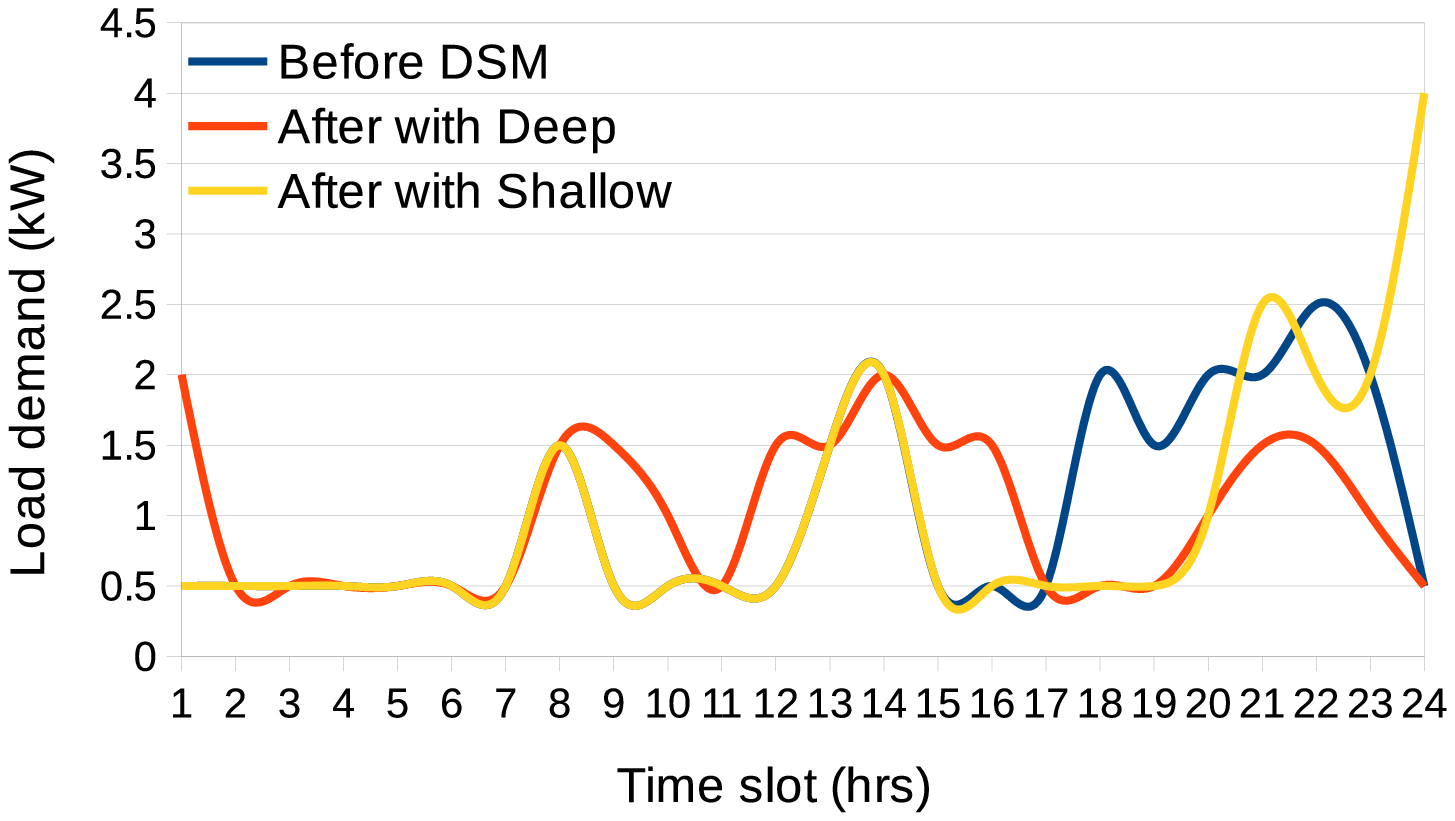}
		\caption{}
	\end{subfigure}%
	\begin{subfigure}{.5\textwidth}
		\centering
		\includegraphics[width=1\linewidth]{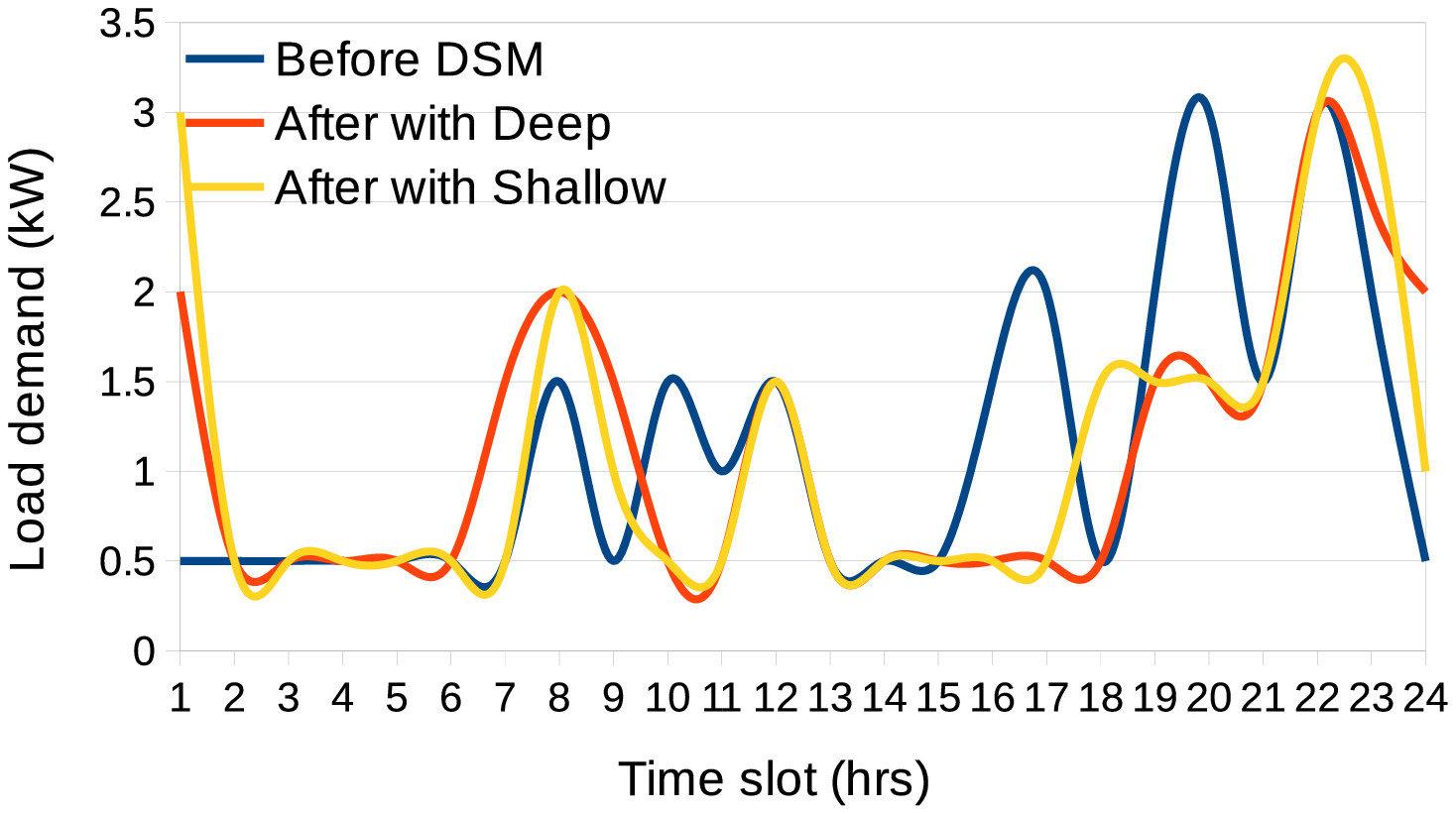}
		\caption{}
	\end{subfigure}
	\begin{subfigure}{.5\textwidth}
		\centering
		\includegraphics[width=1\linewidth]{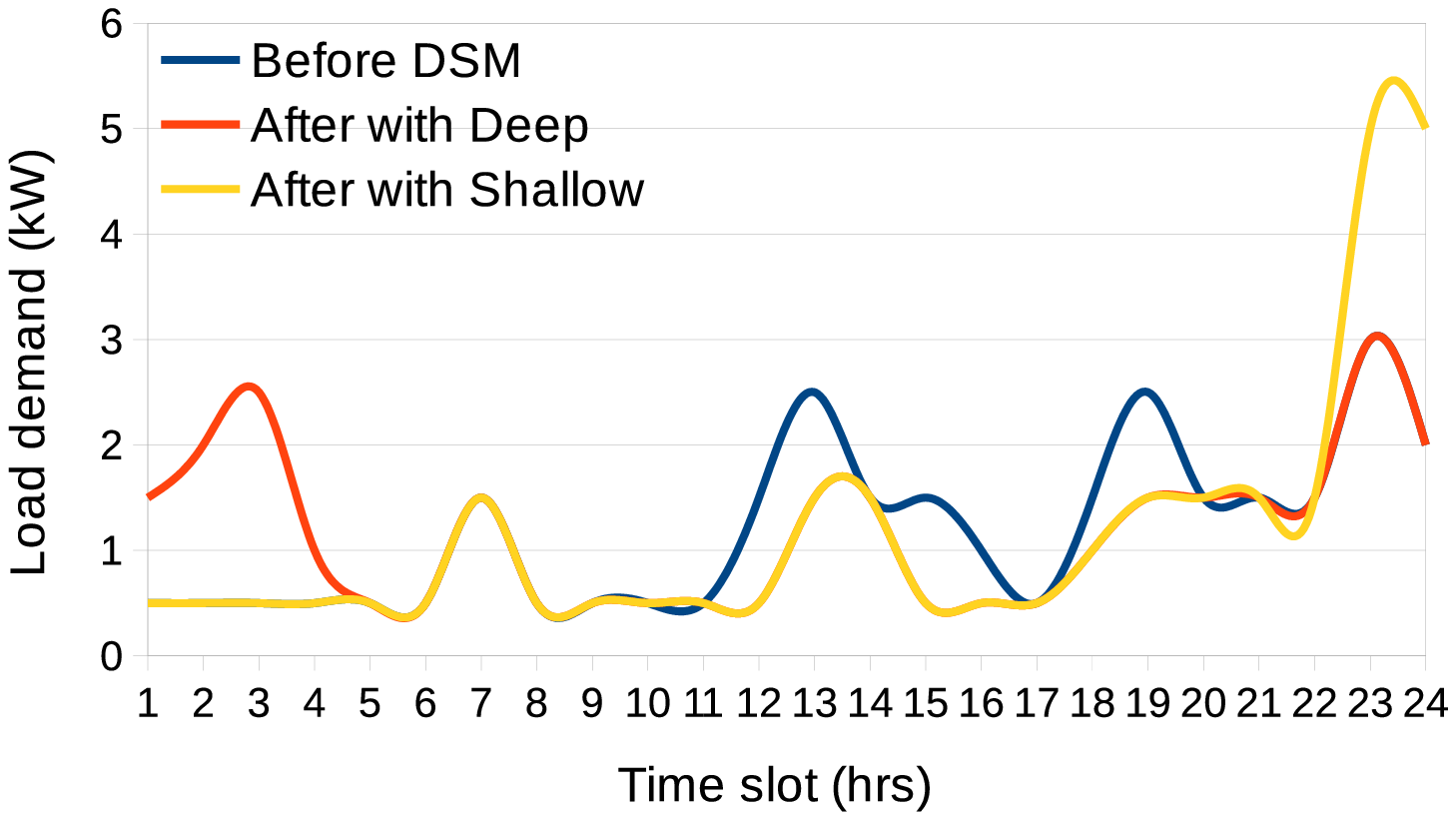}
		\caption{}
	\end{subfigure}%
	\begin{subfigure}{.5\textwidth}
		\centering
		\includegraphics[width=1\linewidth]{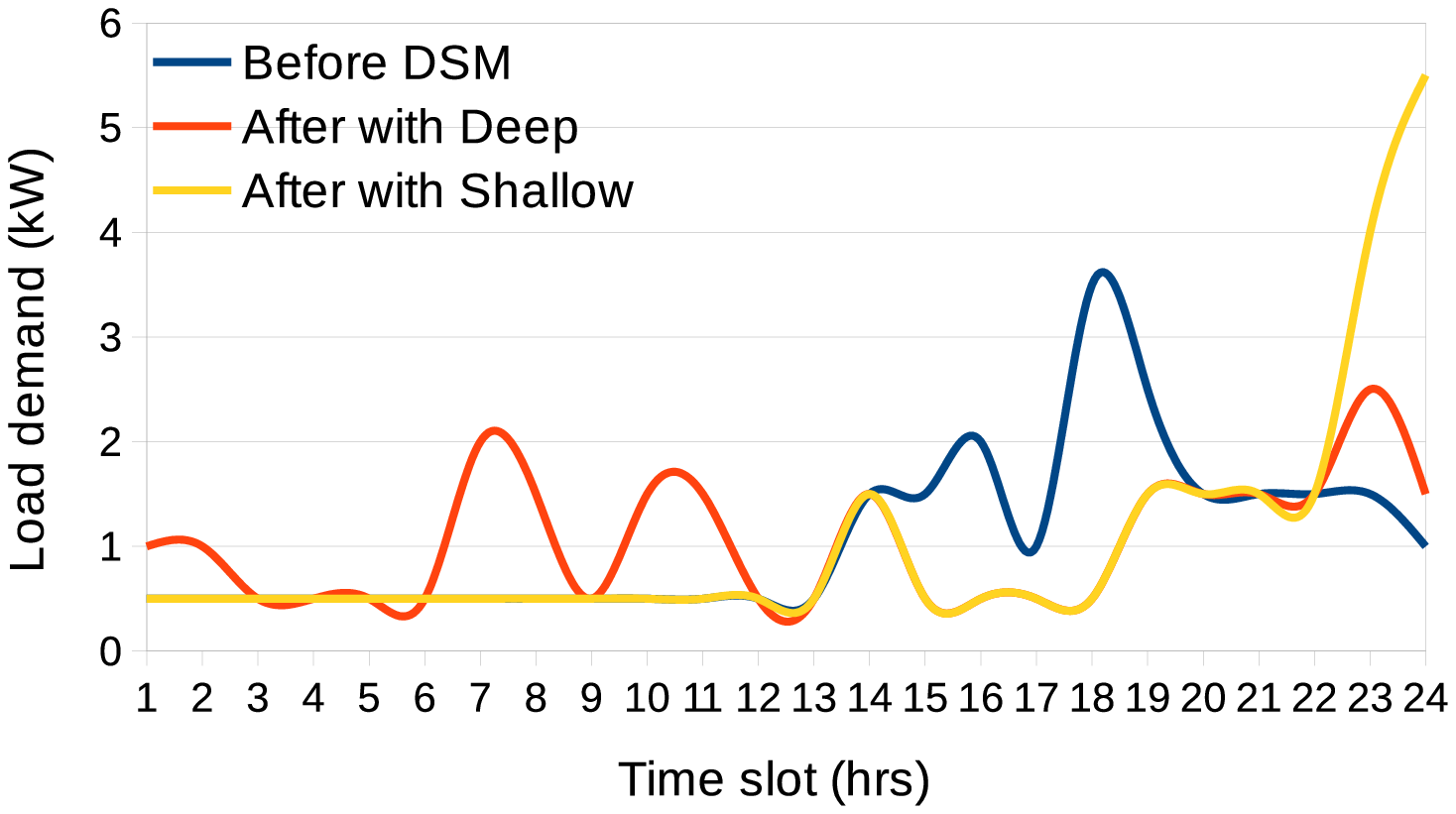}
		\caption{}
	\end{subfigure}
	\begin{subfigure}{.5\textwidth}
		\centering
		\includegraphics[width=1\linewidth]{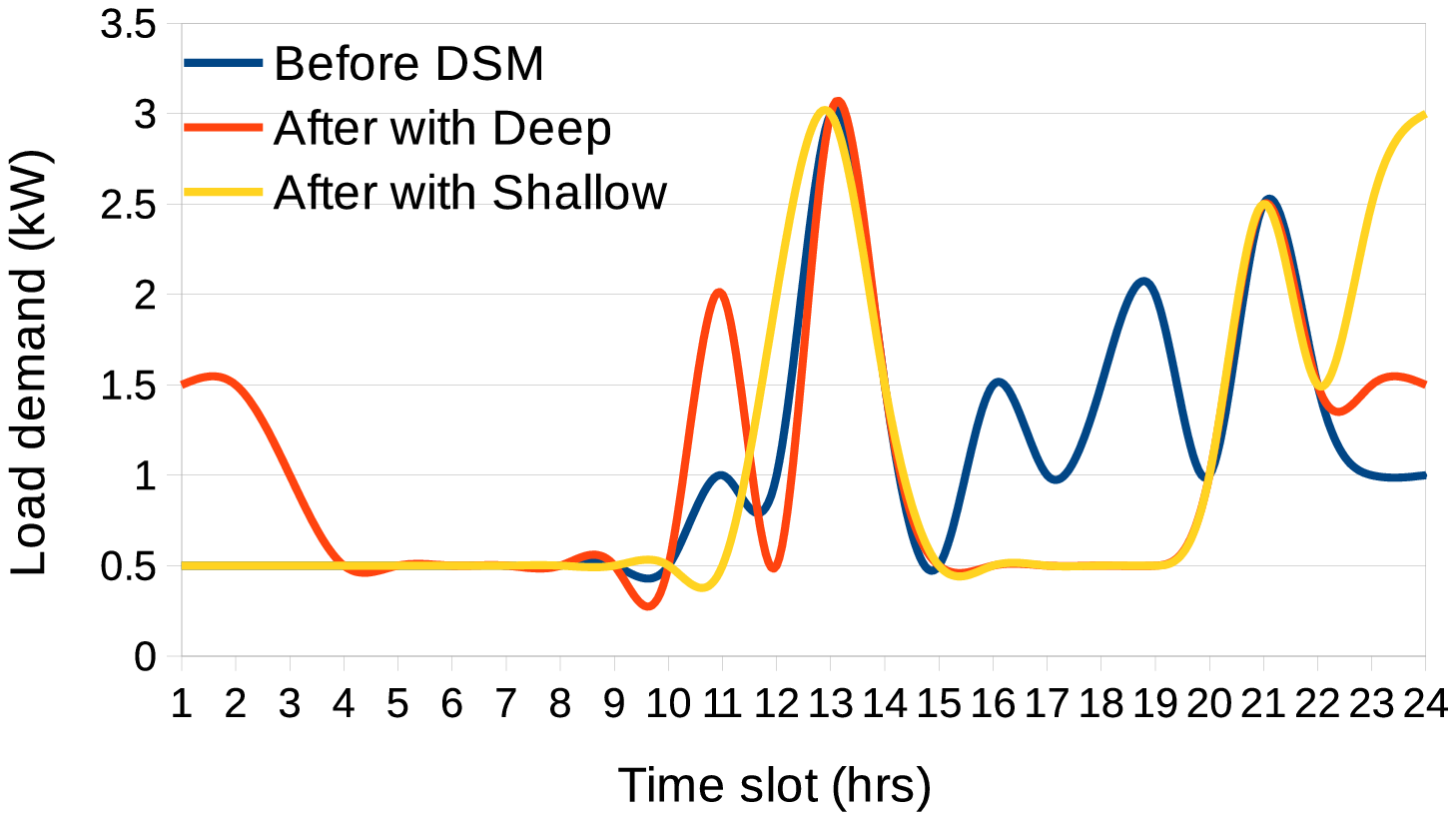}
		\caption{}
	\end{subfigure}
	\begin{subfigure}{.49\textwidth}
		\centering
		\includegraphics[width=1\linewidth]{figs/shallownet/all.eps}
		\caption{}
	\end{subfigure}
	\caption{RL-DSM results\footnotemark[1] of residential loads for five different consumers (a), (b), (c), (d) and (e) respectively with Shallow (After with Shallow) and Deep network (After with Deep). (f) is the RL-DSM results of aggregated residential loads for five different consumers to minimize daily peak load and cost. The figure best viewed in color.}
	\label{fig:shallow}
\end{figure*}

\begin{figure*}[!ht]
	\centering
	\begin{subfigure}{.5\textwidth}
		\centering
		\includegraphics[width=1\linewidth]{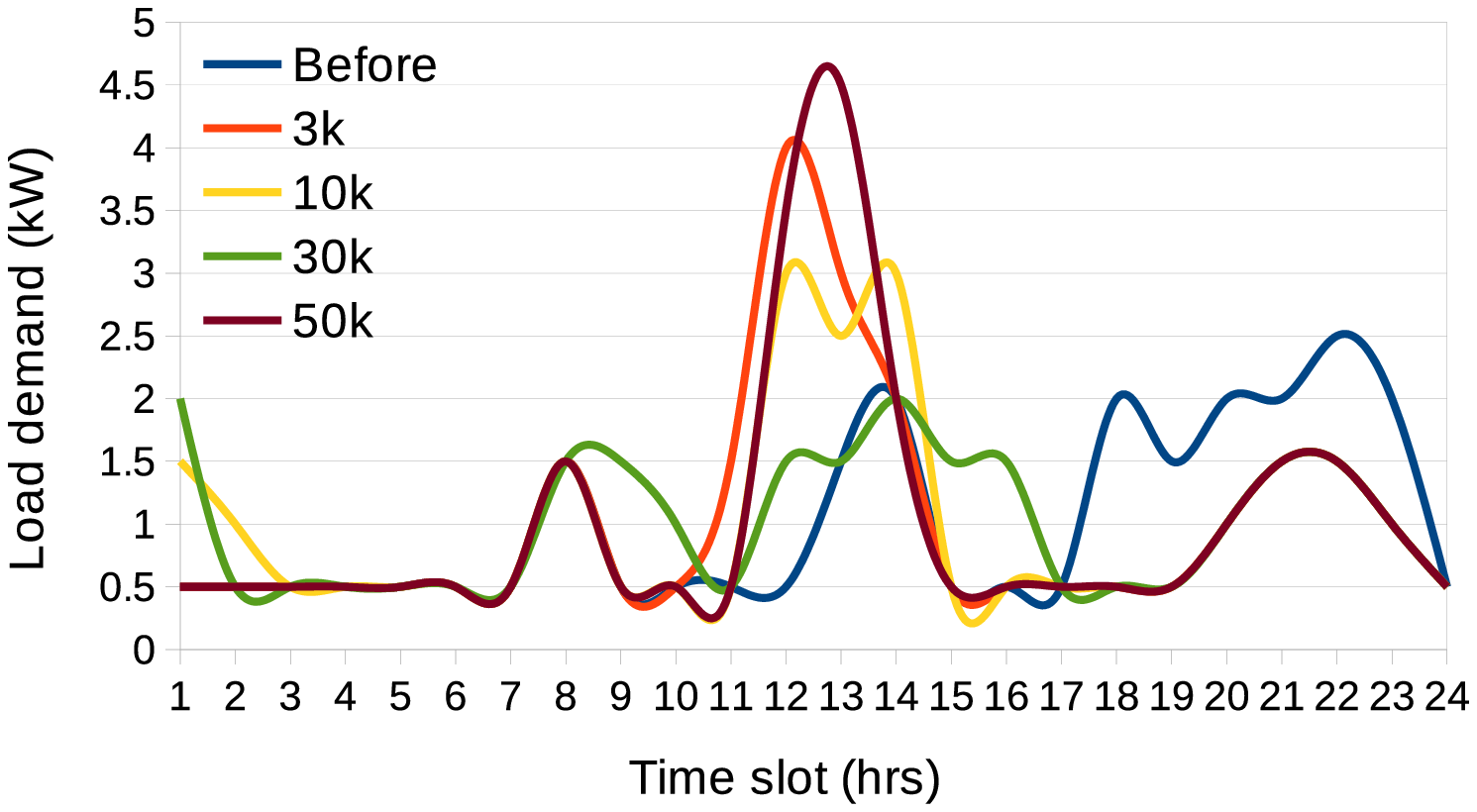}
		\caption{}
	\end{subfigure}%
	\begin{subfigure}{.5\textwidth}
		\centering
		\includegraphics[width=1\linewidth]{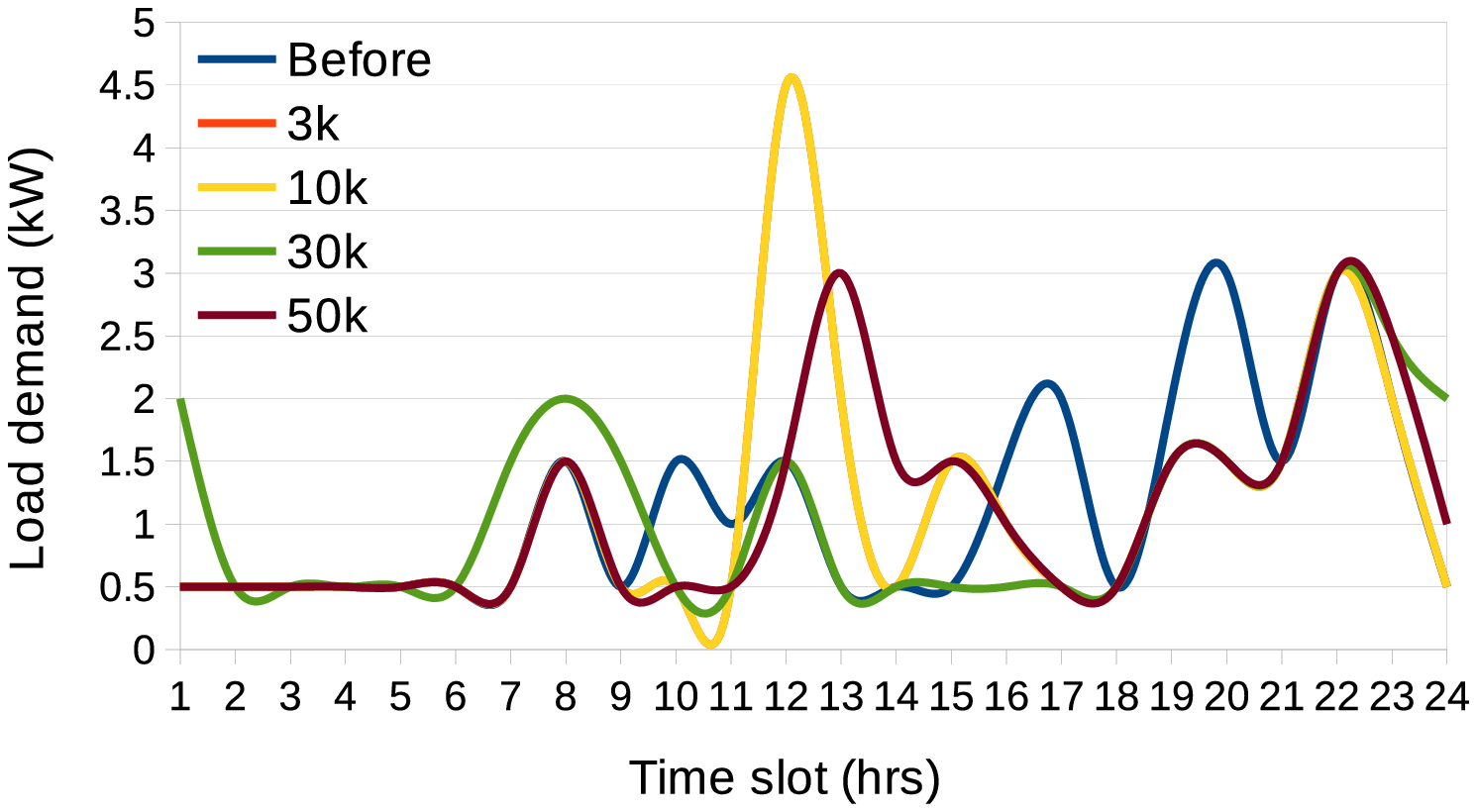}
		\caption{}
	\end{subfigure}
	\begin{subfigure}{.5\textwidth}
		\centering
		\includegraphics[width=1\linewidth]{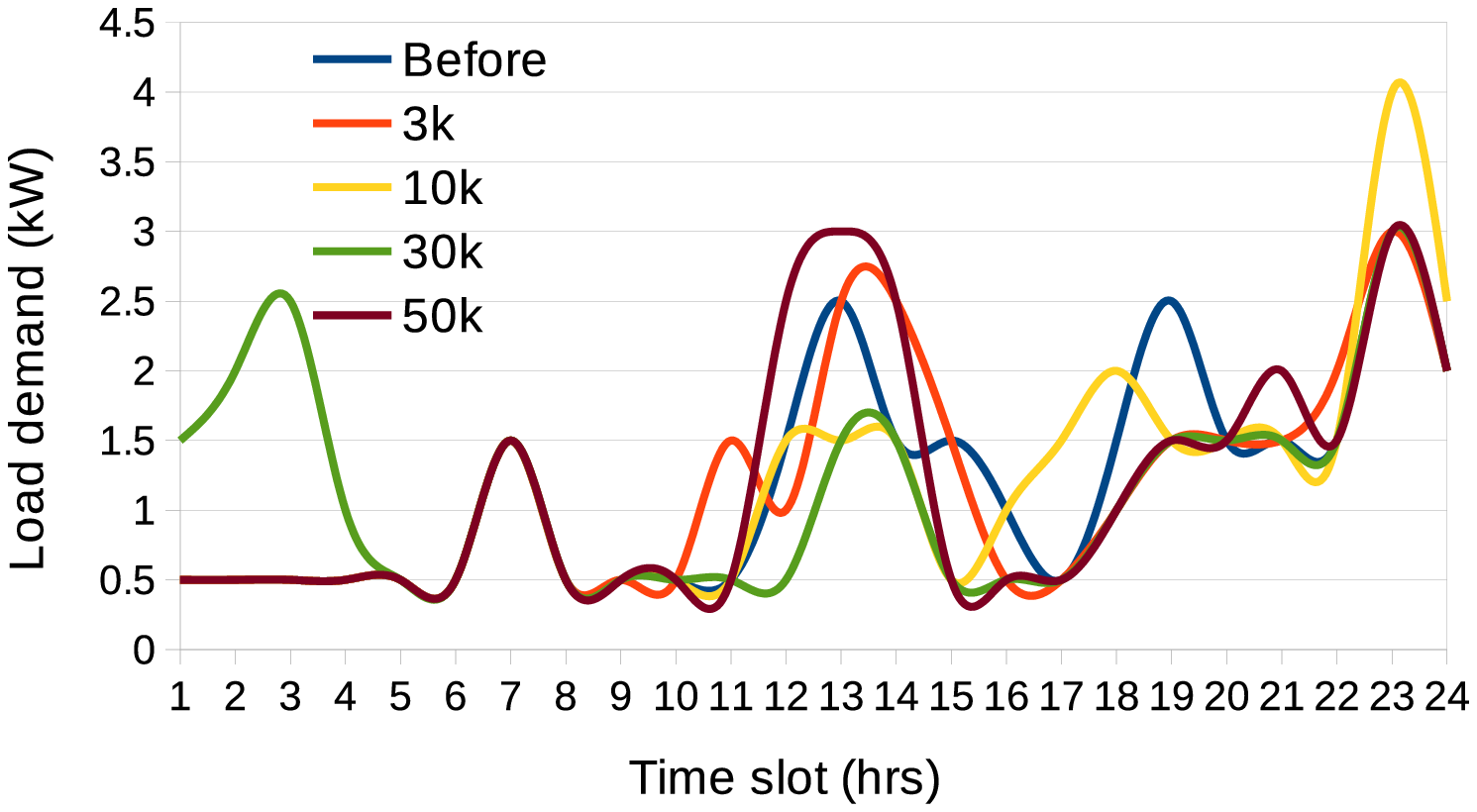}
		\caption{}
	\end{subfigure}%
	\begin{subfigure}{.5\textwidth}
		\centering
		\includegraphics[width=1\linewidth]{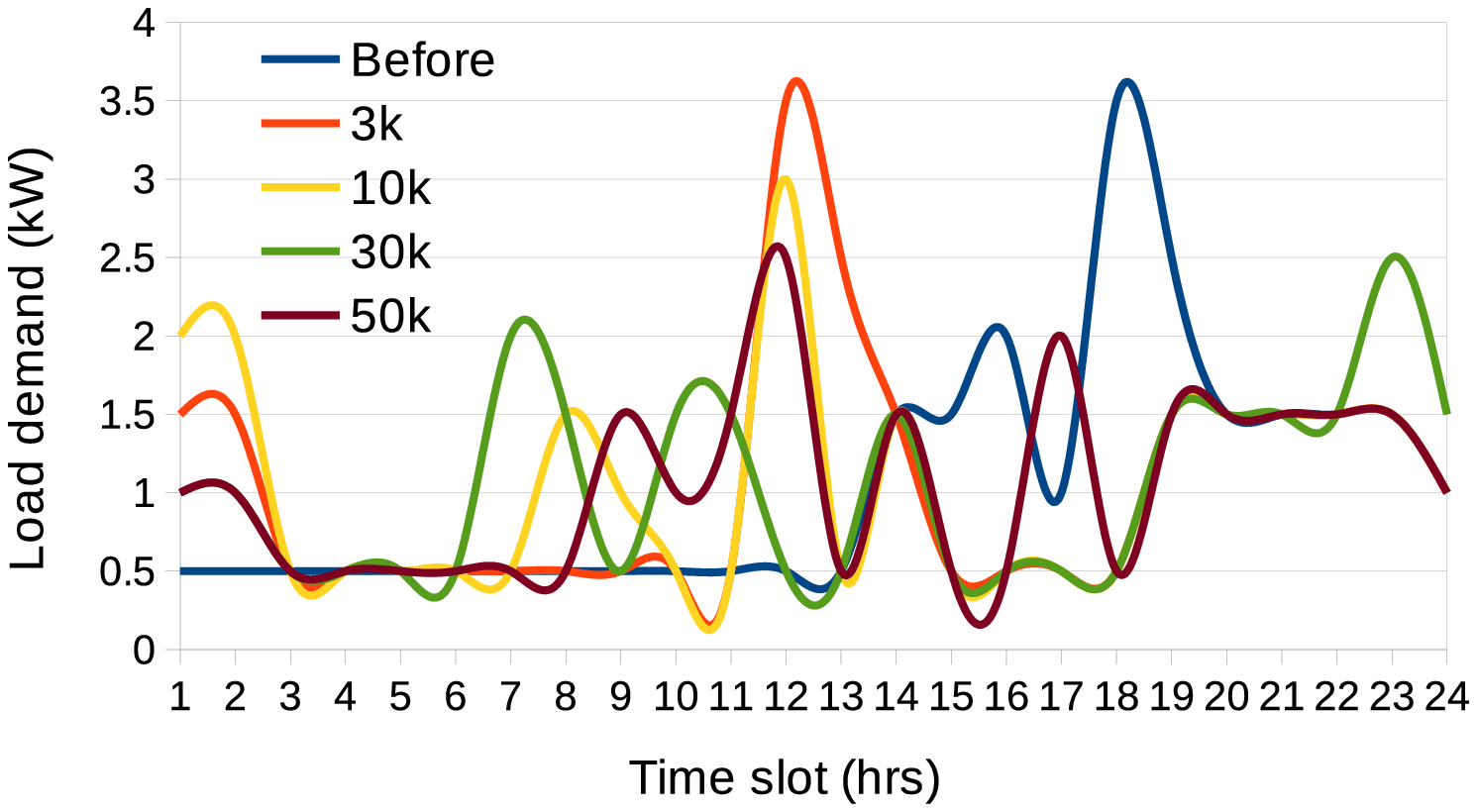}
		\caption{}
	\end{subfigure}
	\begin{subfigure}{.5\textwidth}
		\centering
		\includegraphics[width=1\linewidth]{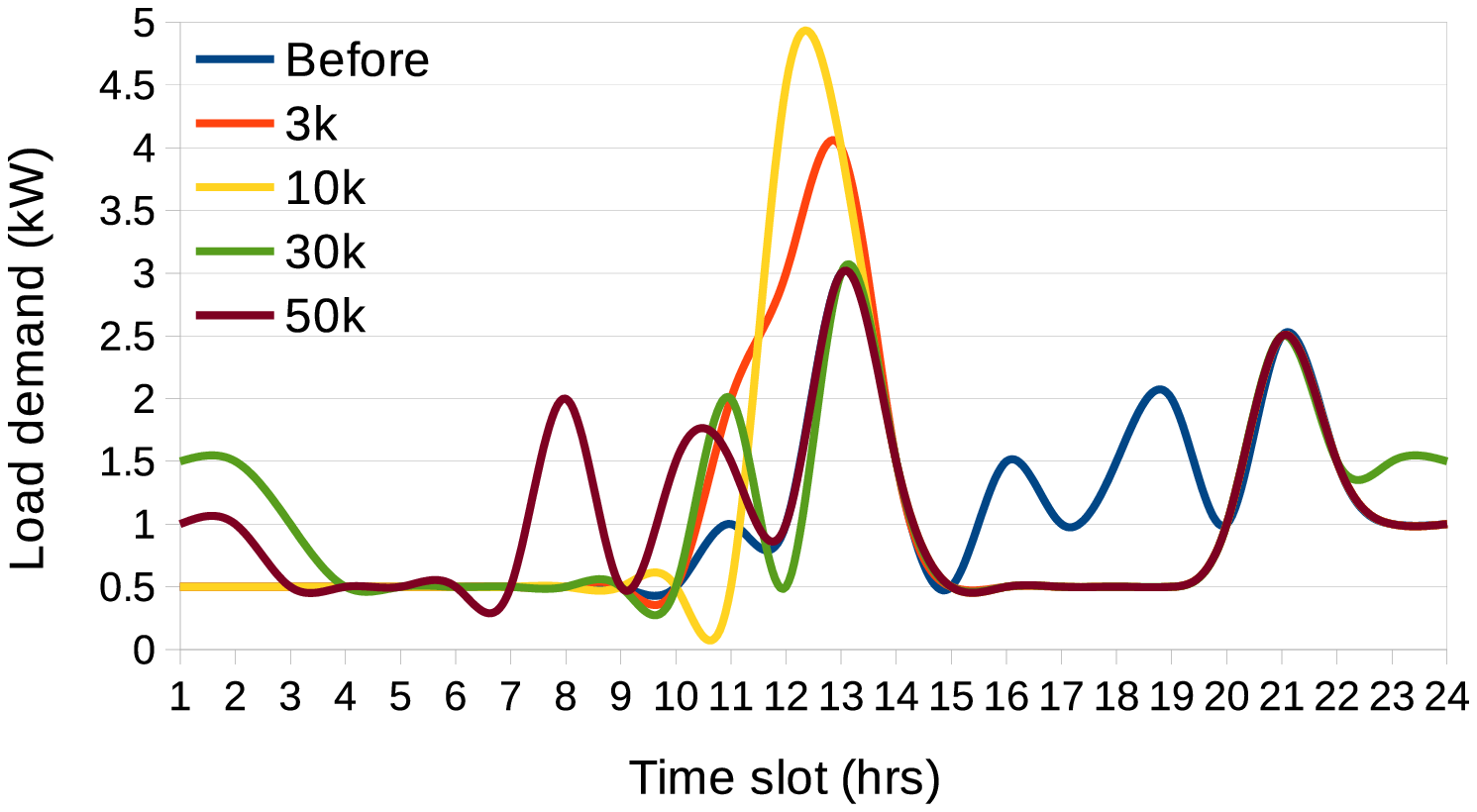}
		\caption{}
	\end{subfigure}
	\begin{subfigure}{.49\textwidth}
		\centering
		\includegraphics[width=1\linewidth]{figs/buffersize/all.eps}
		\caption{}
	\end{subfigure}
	\caption{RL-DSM results\footnotemark[1] of residential loads for five different consumer (a), (b), (c), (d) and (e) respectively using different memory buffer size like 3000(3k), 10000(10k), 30000(20k) and 50000(50k). (f) is the RL-DSM results of aggregated residential loads for five different consumers to minimize daily peak load and cost. The figure best viewed in color.}
	\label{fig:buffersize}
\end{figure*}
	
\end{document}